\documentclass[runningheads]{llncs}

\usepackage[camera-ready,year=2026,ID=1553]{eccv}

\usepackage{eccvabbrv}

\usepackage{graphicx}
\usepackage{booktabs}

\usepackage{amsmath,amsfonts,bm}

\def\eqref#1{equation~\ref{#1}}

\def\1{\bm{1}}

\def\vtheta{{\bm{\theta}}}

\def\vh{{\bm{h}}}

\def\vp{{\bm{p}}}

\def\vz{{\bm{z}}}

\DeclareMathAlphabet{\mathsfit}{\encodingdefault}{\sfdefault}{m}{sl}
\SetMathAlphabet{\mathsfit}{bold}{\encodingdefault}{\sfdefault}{bx}{n}

\newcommand{\softmax}{\mathrm{softmax}}

\usepackage{hyperref}
\usepackage{url}
\usepackage{graphicx}
\usepackage{booktabs}       
\usepackage{amsmath}
\usepackage{multirow}

\usepackage{pgfplots}
\usepackage{makecell}
\usepackage{bm}
\usepackage{comment}
\usepackage{subcaption}
\usepackage{enumitem}
\usepackage{xcolor}

\usepackage[accsupp]{axessibility}  

\usepackage{hyperref}

\usepackage{orcidlink}

\begin{document}

\title{AGE: Adaptive-masking for Graph Embedding in Graph Retrieval-Augmented Generation}

\titlerunning{Abbreviated paper title}

\author{Bao Long Nguyen Huu \inst{1} \and
Atsushi Hashimoto\inst{2}}

\authorrunning{F.~Author et al.}

\institute{OMRON Corporation \and
OMRON SINIC X Corporation}

\maketitle

\begin{abstract}
  GraphRAG is an extension of retrieval-augmented generation (RAG) that supports large language models (LLMs) by referring to graph-structured data as external knowledge.
While this technique ideally captures intricate relationships, it often struggles with graph representations for LLMs, particularly for frozen LLMs, due to the misalignment between graph-based and text-based latent features.
We tackle this issue by introducing the {\it Adaptive-masking for Graph Embedding (AGE)}. AGE employs a Transformer in a mask-based self-supervised learning (SSL) approach. We designed the architecture similar to text embedding encoders, addressing the latent feature misalignment. In contrast to natural language texts, graphs are concise representations, and there exist {\it key nodes} that hold dominant contextual information, which are challenging to predict from their surroundings. Masking such key nodes leads to inefficiency in the SSL process. 
Therefore, AGE focuses on predicting nodes apart from key nodes, utilizing a learnable node sampler.
Our experimental results indicate that AGE significantly improves approaches using non-parametric search component in GraphQA tasks, achieving superior accuracy across four benchmark datasets with distinct characteristics.
  \keywords{Knowledge Graph Question Answering  \and Retrieval-Augmented Generation \and Reinforcement Learning \and Self Supervised Learning.}
\end{abstract}

\section{Introduction}
Large Language Models (LLMs) such as GPT \cite{OpenAI2024,openai_gpt5_2025}, Claude \cite{anthropic_claude_2023}, Gemini \cite{Gemini2023}, Qwen \cite{yang_qwen3_2025}, and LLaMA \cite{llama3.1} have significantly advanced natural language understanding and generation capabilities.
Retriever-Augmented Generation (RAG) \cite{rag_meeting_llms,rag_large_language_models,rag_nlp_survey} integrates query-relevant information into the generation process, enabling LLMs to access and utilize domain-specific knowledge beyond their pretraining corpus.
However, although RAG enhances LLMs with external data, it may struggle to capture essential structured relationships, reducing search precision and reasoning effectiveness \cite{zeng2024perceive,yao2024tree}.
\begin{figure*}[]
\begin{center}
\includegraphics[width=0.98\linewidth]{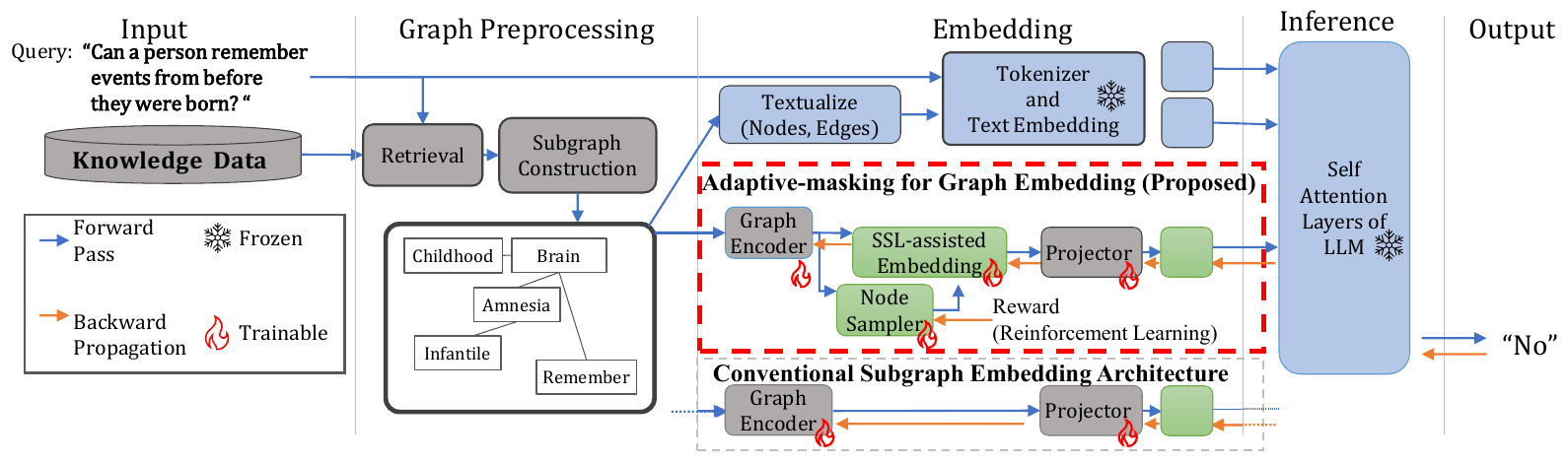}
\end{center}
\vspace{-3mm}
   \caption{
   Overview of GraphRAG with the proposed {\it Adaptive-masking for Graph Embedding} (AGE) embedding.
   1) Retrieval: Find graph elements relevant to the query using a non-parametric process.
   2) Subgraph Construction: Extend retrieved graph elements with their adjacencies \cite{G-Retriever}.
   3) Embedding: Use tokenizer and text embedder for textualized graph and query. Apply AGE for structured relationships of the graph.
   4) Inference: Input embeddings into LLM to generate an answer.}
\label{eyecatch}
\end{figure*}
Graph Retriever Augment Generation (GraphRAG) \cite{graph_rag_query_focused_summarization,gnn_rag_graph_neural_retrieval} is a technology that uses graphs to overcome the limitations of RAG. 
Graph data, represented by nodes (entities) and edges (relationships), clearly presents complex relationships. 
This provides several benefits, such as facilitating data integration \cite{grag_graph_retrieval_augmented_generation}, improving search accuracy \cite{rag_vs_graphrag,lego_graphrag}, enhancing inference capabilities \cite{guo2025empowering,han2025retrieval}, and reducing hallucinations \cite{G-Retriever}.
By capturing sub-graphs, the broader context and interconnections within the graph structure can be captured, enabling comprehensive information to be accessed for LLMs enhance the performance in domain-specific tasks.
\\
This study investigates GraphRAG methods that operate within practical computational costs. Fine-tuning LLMs can enhance GraphRAG performance, yet it is resource-intensive. Instead, previous methods often focused on the retrieval module as it is a key factor for GraphRAG performance. Trainable retrievers, such as LLM-based retrievers \cite{think_on_graph,reasoning_on_graphs} realize a higher retrieval accuracy. However, this strategy still requires significant computational overhead. Non-parametric retrievers \cite{G-Retriever,qa_gnn} are efficient and low-cost but may contain redundant or missing critical nodes, leading to the lack of explicit structural constraints. To maintain practicality, we base our method on non-parametric retrievers with frozen LLM, and improve performance of structural representation by updating the graph embedding module.
Embeddings play a crucial role in bridging the gap between retrieved graph data and the LLM input space. Multiple methods \cite{G-Retriever, Graphtoken} use graph embedding together with textualized graph representation (Fig. \ref{eyecatch}), indicating that the graph encoder should embed relationships between elements, rather than their individual contents. 
What is the optimal strategy to achieve such encoding for a frozen LLM?
We considered two factors: similarity of the embedding space to the LLM's text encoder and its relationship embedding capability.
Since the LLM's text encoder uses mask-based SSL  \cite{liu2019roberta,reimers2019sentence,openai2022embedding}, which learns to embed relationships between elements into the embedding space by optimize the reconstruction of masked elements.
These factors aim to embed relationships between nodes within the retrieved subgraph into the embedding space by adapting the mask-based SSL with minimal modification.
\\
To realize this intention, we propose a novel embedding strategy, the {\it Adaptive-masking for Graph Embedding (AGE)}. The architecture of AGE is designed to imitate the general self-supervised text embedding process, while incorporating the Joint-Embedding Predictive Architecture (JEPA) \cite{LeCun2022APT}, which improves representations in the embedding space by eliminating unnecessary detail reconstruction. Although generative SSL shows promise, the quality of reconstruction relies on the discriminability of input nodes \cite{Wei2022Masked,Chien2022Node}. Random masking fails on non-discriminative nodes, leading to poor representations \cite{bizeul2024masking,seong2025rethinking}.
To avoid this, the only major modification is adding a node sampler trained via reinforcement learning (RL) to selectively mask nodes, replacing traditional random node masking with an adaptive approach. The motivation for this approach stems from the fact that, graphs are concise logical structures with minimal redundancy. Hence, some nodes are crucial for maintaining graph integrity; we refer to them as {\it key nodes}. Our RL-based strategy aims to guide SSL to distinguish the representations of key and auxiliary nodes, encouraging LLMs to identify redundant information within the retrieved graph.
The contributions of this paper are outlined as follows:
\begin{enumerate}
    \item  We propose AGE, a novel method that represents retrieved subgraphs via key-node and auxiliary-node embedded by RL-guided mask-based SSL.
    \item Our study reveals adaptive masking approach's notable effectiveness over random masking within GraphRAG.
    \item AGE uses a non-parametric retriever and open LLMs, while also achieving SOTA on three other benchmarks.
\end{enumerate}
%
%
%
%
%
\vspace{-3mm}
\section{Related Work}
\subsection{Graph Representation for LLMs} 
In the context of representing graphs as input to LLMs, it is necessary to first convert the retrieved graph data into specific formats. We summarize two distinct formats: textualization and graph embeddings. Textualization \cite{fatemi2024talk,graph_chain_of_thought,li2024enhanced} is a text-based formalization method designed to characterize and represent graph data. Node sequences are a popular form of textualization\cite{chen2024llaga,reasoning_on_graphs,think_on_graph}. Some methods \cite{reasoning_on_graphs,think_on_graph,plan_on_graph} propose LLM-based retrievers to extract reasoning paths. A node sequence ordered along the path aids LLM's reasoning. However, many studies report negative conclusions in interpreting text-encoded graphs with concurrent LLMs \cite{huang2023can,guo2023gpt4graph,wang2023can}, suggesting a need for solutions beyond textualization.
\\
The other format, graph embeddings, have recently been adopted in GraphToken \cite{Graphtoken}. Following this, G-Retriever \cite{G-Retriever} proposed a retrieval framework for graph embeddings. This occurs when graph embeddings are added as tunable prompts to the LLM in addition to their textualized representations. 
In this work, we improve the quality of LLM responses on the G-Retriever framework through enhancing the representation of graph embeddings.
Some methods \cite{xu2025amar,grag_graph_retrieval_augmented_generation} build self-alignment and cross-question module among retrieved entities, relations, and subgraph embedding elements. Some methods enhance embeddings through a two-stage training process \cite{ji2024ntllm,wang2024llmsas}. The first stage trains the embedding module on SSL alone; in the second stage, prompt tuning aligns the structured relationships embedded for LLM input by the pretrained module. 
Since each LLM has its own domain embeddings and input spaces, two-stage training process prioritize maximizing performance. Instead, focusing on practical, we propose a one-stage training process SSL that integrates with prompt tuning.

\subsection{Self-Supervised Learning} 
Many existing self-supervised learning architectures focus on learning representations that effectively capture relationships between input data.
Joint-Embedding Architecture (JEA) \cite{bardes2021vicreg,caron2020unsupervised,grill2020bootstrap} has shown considerable promise in advancing SSL methodologies. Joint-Embedding SSL for GNNs, such as GraphCL \cite{ying2021do}, GCA \cite{zhu2021graph} and JOAO \cite{you2021graph}, learn node representations by contrasting positive and negative samples. 
Subsequent studies identified areas for enhancement in JEA \cite{chin2024masking,jing2021understanding,lee2025theoretical}, particularly the issue of mapping all inputs to a single constant vector, known as the collapsing problem.
Generative Architecture (GA) \cite{MaskedAutoencoders2021,baevski2022data2vec,devlin2018bert} focuses on reconstructing masked portions of the input at either the pixel or token level.
GraphMAE \cite{hou2022graphmae,hou2023graphmae2} learns representations by reconstructing masked samples. These methods encourage the model to learn more robust and diverse representations, potentially reducing the risk of representation collapse \cite{chin2024masking,jing2021understanding,I-Jepa}. 
Joint-Embedding Predictive Architectur (JEPA) \cite{LeCun2022APT} eliminate reconstruction of pixel or token-level details and enhances the semantic level of self-supervised representations \cite{I-Jepa,V-Jepa}. In this work, we first demonstrate JEPA's effectiveness in the GraphRAG framework. JEPA originates from cognitive neuroscience, suggesting that humans have an ability of top-down schema reasoning, aiding planning, decision-making, and problem-solving on complex tasks \cite{tang2007top,mittal2020learning,theves2021learning}.
In GraphRAG, the tokens fed into the LLM's hidden layer should capture the ability. We implement it as JEPA in the graph embedding module. Cognitive science has revealed that a brain region called the temporal lobe plays a role in bottom-up associative learning, selecting key knowledge and linking it to related data \cite{jackson2018emergent,edmonds2019decomposing,cox2024representational}. These selection processes can also occur with new information. Aiming to reproduce this functionality in GraphRAG, AGE has the novel node sampler module.
%
%
%
%
%
\vspace{-3mm}
\section{Preliminaries}
\vspace{-3mm}
\textbf{GQA with LLM.} For a query $q$ on a textual graph $G$, there is an optimal subgraph $\overline{S^{*}} \in S(G)$ and query relevant text-modal knowledge $T^{*}$ that guides the LLM to produce expected answers, where $S(G)$ is the set of all subgraphs of $G$. The challenge of GraphRAG is to efficiently search for the relevant subgraph $S^{*}$ and represent it to $\overline{S^{*}}$ for an LLM $p_\Phi$ improve generation. The probability distribution of the output sequence $Y$ is given by:
\begin{equation}
p_\Phi(Y \mid [q, G]) =  \prod_{i=1}^{n} p_\Phi(y_i \mid y_{<i}, [q, T^{*}, \overline{S^{*}}]),
\end{equation}
where $y_{<i}$ represents the prefix tokens, and $[q, \overline{S^{*}}]$ indicates the concatenation of the query, relevant text-modal knowledge and optimal subgraph information, respectively.
\\
\textbf{Joint-Embedding Predictive Architecture.}
Mask-based SSL methods such as MAE are well suited to handle corrupted input, as they learn to reconstruct missing or corrupted input parts. JEPA improves upon MAE by eliminating the reconstruction of unnecessary input feature details, focusing instead on learning more abstract representations. 
JEPA consists of an encoder $E_\theta(\cdot)$, predictor $P_{\phi}(\cdot)$ and target encoder $T_{\theta}(\cdot)$. The stop-gradient operation $\operatorname{sg}$ is employed to prevent representation collapse in the target encoder $T_{\theta}(\cdot)$. The predictor generates $y$ from visible input $x$ and masked input $\Delta_x$. The encoder and predictor are trained simultaneously with the objective:
\begin{equation}
\min || P_{\phi}\big(\Delta_x, E_{\theta}(x)\big) - \operatorname{sg}\big(T_{\theta}(y)\big) ||_2,
\end{equation}
The loss is applied only to the predictions of the masked input $\Delta_x$.
\\
\textbf{Reinforcement Learning.}
To estimate the key and auxiliary nodes on retrieved graph for mask-based SSL discriminative embedding.
We adopt REINFORCE \cite{Sutton2000Policy}, a basic policy gradient method in RL. Let $\mathcal{D}={(q,Y^{*})}$ denote a corpus of training data, where $Y^{*}$  is the complete reference label for query $q$. REINFORCE optimizes a policy $\pi_{\theta}$ parameterized by $\theta$, to maximize reconstruction quality $R_a$ for each masking action $a$. The policy gradient is given by:
\begin{equation}
\nabla_\theta \mathcal{J}(\theta) = \mathbb{E}_{q \sim \mathcal{D}, S^{*} \sim q} \left[ \sum_{v \in S^{*}} \nabla \log \pi_\theta(a | v) \cdot R_a \right]
\end{equation}
where where $\mathcal{J}(\theta)$ is the expected return. $\pi_\theta(a | v)$ denotes the probability distribution estimated by policy $\pi_{\theta}$ for taking action $a$ (masking or not) on node $v$. The SSL framework with masked node reconstruction serves as the RL environment.
%
%
%
%
%
%
\vspace{-3mm}
\section{Approach}
\vspace{-3mm}
Our framework, illustrated in Figure \ref{eyecatch}, consists of four main steps: {\it input}, {\it graph preprocessing}, {\it embedding}, and {\it inference}. 
We adopt the previous method \cite{G-Retriever} that applies SentenceBert \cite{reimers2019sentence} to indexed knowledge data at ({\it input}) step and employ a static k-nearest neighbors \cite{Kramer2013kNN} retrieval approach combined with Prize-Collecting Steiner Tree \cite{Bienstock1993PrizeCollectingTSP} subgraph construction during {\it graph preprocessing}. For {\it inference}, we can use arbitrary LLMs, as usual RAG methods. Therefore, this section focuses on the details of {\it embedding} step.
%
\subsection{Text Embedding of Query and Text Graph}
We transform the retrieved subgraph $S^{*}$ into a textual format, following \cite{G-Retriever} as in the first two steps.
The converted text is then concatenated with the input query $q$. The concatenated texts are embedded into $h_{\text{text}}$ using a pretrained function for the frozen LLM, $\text{TextEmbedding}$, where $[;]$ denotes concatenation and $L$ is the output token sequence length, as follows:
\begin{gather}
  h_{\text{text}}=\text{TextEmbedding}(\text{[textualize}(S^{*});q]) \in \mathbb{R}^{L\times d_{l}},
\end{gather}
%
%
\newcommand{\myvh}[1]{\vh_{\text{#1}}}
\newcommand{\vhin}{\myvh{in}}
\newcommand{\vhout}{\myvh{out}}
\newcommand{\vhkey}{\myvh{key}}
\newcommand{\vhtarget}{\myvh{target}}
\newcommand{\myvz}[1]{\vz_{\text{#1}}}
\newcommand{\vzms}{\myvz{NS}}
\newcommand{\vzkey}{\myvz{key}}
\newcommand{\vzaux}{\myvz{aux}}
\newcommand{\param}{\vtheta}
\begin{figure*}[!tb]
\begin{center}
\includegraphics[width=0.95\linewidth]{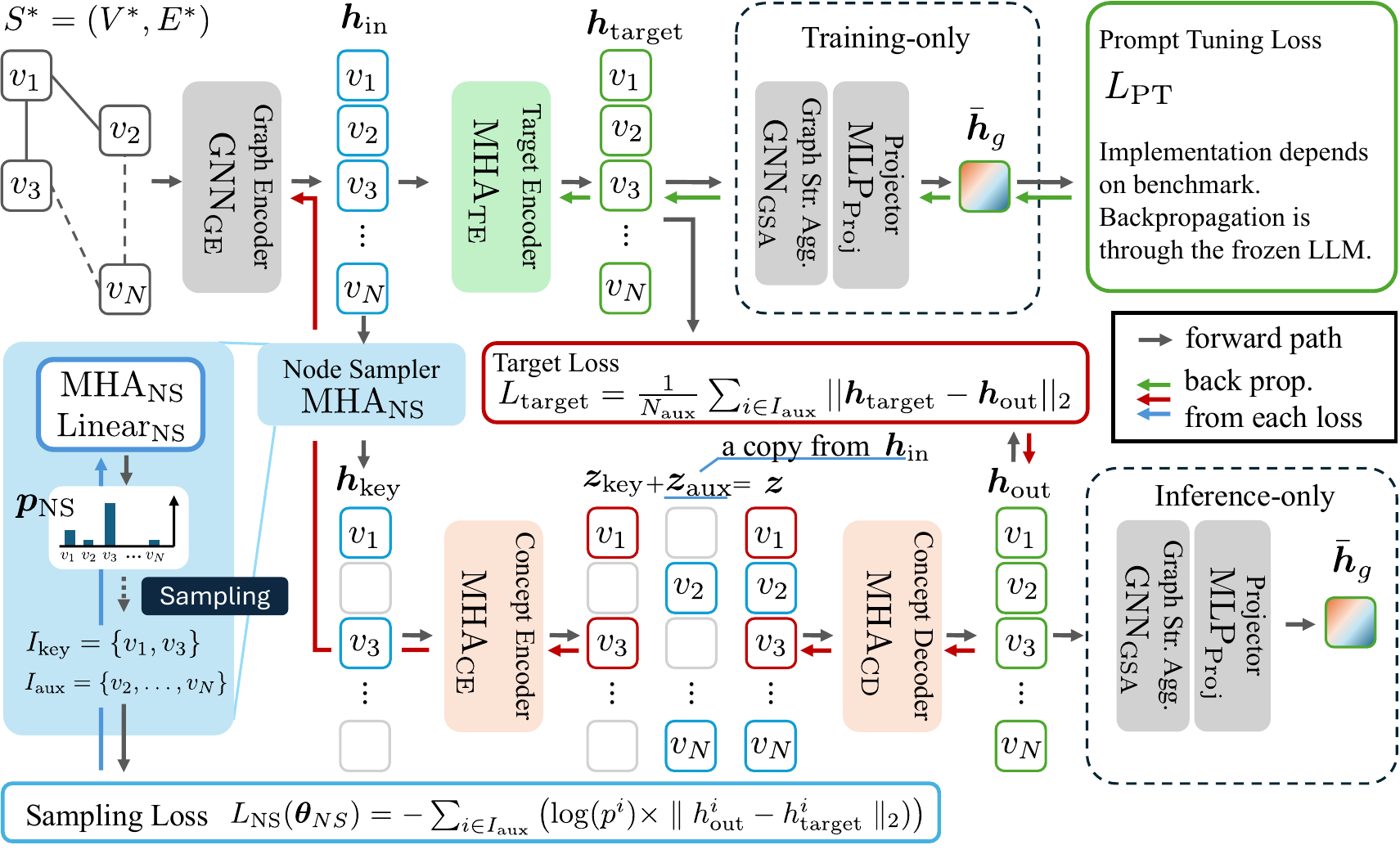}
\end{center}
\vspace{-3mm}
   \caption{Architecture for Adaptive-masking for Graph Embedding: During training, $\vhtarget$ is connected to the downstream for the target encoder training, while $\vhout$ is used during inference. The node sampler explores the optimal distribution for mask-based SSL for graphs. The loss functions train distinct sets of modules without overlap.}
\label{AGE}
\end{figure*}
\vspace{-3mm}
\subsection{Adaptive-masking for Graph Embedding}
\vspace{-3mm}
AGE comprises a {\it node sampler}, {\it concept encoder-decoder}, {\it target encoder}, and {\it graph-structure-based aggregator} modules, as overviewed in Fig. \ref{AGE}. The AGE's input, $\vhin$, is encoded from the retrieved graph $S^{*}$ with a conventional graph encoder. $\vhin$ is then passed to {\it node sampler} and {\it target encoder}.
The node sampler categorizes the nodes into {\it key nodes} and the remaining {\it auxiliary nodes}.
The selected {\it key nodes} are fed to {\it concept encoder-decoder}.
The output $\vhout$ is trained to predict
 $\vhtarget$, the output of {\it target encoder}, forming a JEPA.
The embedding $\vhout$ is then aggregated into a token to be fed to the LLM.
The rest of this subsection explains each module in Fig. \ref{AGE} individually.
\newcommand{\GNNge}{\text{GNN}_{\text{GE}}}
\newcommand{\paramge}{\param_{\text{GE}}}
\\
\textbf{Graph Encoder} prepares input for AGE.
The retrieved subgraph $S^{*}=(V^{*},E^{*})$ consists of query-relevant nodes $V^{*}$ and edges $E^{*}$. 
$\vhin$ is obtained from $S^{*}$ by the graph encoder $\GNNge$ as follows:
\begin{gather}   
\vhin=\GNNge(S^{*};\paramge)\in \mathbb{R}^{N\times d_{g}},  
\end{gather}   
where $\paramge$ is parameter of $\GNNge$, $d_{g}$ is dimension of each output node feature, and $N=|V^{*}|$.
\newcommand{\prms}{\vp_{\text{NS}}}
\newcommand{\MHAms}{\text{MHA}_{\text{NS}}}
\newcommand{\paramms}{\param_{\text{NS}}}
\newcommand{\Nkey}{N_{\text{key}}}
\newcommand{\Naux}{N_{\text{aux}}}
\newline
\textbf{Node sampler} estimates key nodes using $\vhin$ from the graph encoder for adapting masks on auxiliary nodes.
The node sampler processes $\vhin$ through a Multi-Head Attention (MHA) network, a linear layer, and a softmax activation, to obtain nodes probability scores $\prms$ as follows:
\begin{gather}   
\vzms=\MHAms(\vhin;\paramms^{\text{MHA}}) \in \mathbb{R}^{N \times d_{g}}, \\
\prms=\text{Softmax}(\text{Linear}(\vzms;\paramms^{\text{Linear}})) \in [0,1]^{N\times 1},
\label{eq:node_sampler}
\end{gather}
where $\paramms=\{\paramms^{\text{MHA}},\paramms^{\text{Linear}}\}$ is the parameter of this module.
We sample $\Nkey$ nodes based on a categorical distribution defined by $\prms$, where we decide $\Nkey$ by the sampling rate $\rho$ as $\Nkey=\lceil \rho N\rceil$.
Hereafter, we denote the sampled key nodes as $I_{\text{key}}$ and auxiliary nodes as $I_{\text{aux}}(=V^{*}\backslash I_{\text{key}})$.
Based on $I_{\text{key}}$, we extract key node features $\vhkey \in \mathbb{R}^{k \times d_{g}}$ from $\vhin$ and input it to our concept encoder-decoder module.
\newcommand{\MHAce}{\text{MHA}_{\text{CE}}}
\newcommand{\MHAcd}{\text{MHA}_{\text{CD}}}
\newcommand{\paramce}{\param_{\text{CE}}}
\newcommand{\paramcd}{\param_{\text{CD}}}
\newcommand{\paramced}{\param_{\text{CED}}}
\\
\textbf{Concept Encoder-Decoder} consists of a concept encoder $\MHAce$ and a concept decoder $\MHAcd$.
$\MHAce$ encode the input $\vhkey$ into the latent representation $\vzkey$ as follows:
\begin{gather}   
\vzkey=\MHAce(\vhkey + \text{PE}(\vhkey);\paramce) \in \mathbb{R}^{k \times d_{g}}, \ 
\end{gather}
where $\text{PE}(\cdot)$ represents positional encoding as defined by \cite{ma2021graphattentionnetworkspositional}, and $\paramce$ is the parameter of $\MHAce$.
%
%
$\vzkey$ is combined with $\vzaux$, placeholder vectors for unsampled auxiliary nodes with values copied from $\vhin$, as in \cite{zheng2025exlm}.
Let $\vz \in \mathbb{R}^{N\times d_g}$ be the combined node features, which maintains $\vhin$'s original node position.
$\MHAcd$ decodes $\vhout$ with the positional encoding $\text{PE}$ as follows:
\begin{gather}   
\vhout=\MHAcd(\vz+\text{PE}(\vz);\paramcd) \in \mathbb{R}^{N \times d_{g}},
\end{gather}
where $\paramcd$ is the parameter of $\MHAcd$. $\vhout$ is the output of AGE, which we train to predict $\vhtarget$, an embedding obtained from all nodes through the target encoder.
%
\newcommand{\MHAte}{\text{MHA}_{\text{TE}}}
\newcommand{\paramte}{\param_{\text{TE}}}
\\
\textbf{Target Encoder} is applied to obtain a prediction target for the previous module in a semantic space (JEPA), as it works more robustly than in the input space (GA) \cite{I-Jepa,chen2025denoising, fei2023ajepa, V-Jepa}.
The target encoder $\MHAte$ projects $\vhin$ to a target embedding $\vhtarget$ as follows: 
\begin{gather}   
\vhtarget=\MHAte(\vhin+\text{PE}(\vhin);\paramte) \in \mathbb{R}^{N \times d_{g}},
\end{gather}
where $\paramte$ is the parameter of $\MHAte$.
We train $\MHAte$ with downstream tasks, optimizing it to produce embeddings that directly contribute to the task.
$\MHAce$ and $\MHAcd$ are trained in parallel with $\MHAte$, with the target encoder learning graph representations for LLMs and the concept encoder-decoder learn to exploit key-node representations for mimic the target encoder's auxiliary node representations.
Inferring masked auxiliary nodes condenses relational concepts between key and auxiliary nodes into $\vzkey$ (and thus $\vhout$ in the downstream).
This JEPA-derived mechanism would be synergetic with GraphRAG as the previous studies suffers from embedding graph's structured relationships efficiently \cite{huang2023can}.
%
%
%
\newcommand{\GNNgd}{\text{GNN}_{\text{GSA}}}
\newcommand{\paramgd}{\param_{\text{GSA}}}
\newcommand{\MLPproj}
{\text{MLP}_{\text{Proj}}}
\newcommand{\paramproj}{\param_{\text{Proj}}}
\\
\textbf{Graph-structure-based Aggregator} $\GNNgd$ projects $\vhout$ to a single token $\bar{\vh}_{g}$. As the graph encoder module, we followed \cite{G-Retriever} for this module. 
It aggregates $\vhout$ referring $E^{*}$, the edge connections of the original subgraph $S^{*}$. Where $\text{POOL}$ is mean pooling and $d_{l}$ is the input dimension of the target layer.
The projector $\MLPproj$ adjusts the aggregated embeddings to fit the LLM's input dimension.
\begin{gather} 
\vh_{g} = \text{POOL}(\GNNgd(\vhout;\paramgd)) \in \mathbb{R}^{d_{g}}, 
\; \; \; \;
\bar{\vh}_{g} = \MLPproj(\vh_{g};\paramproj) \in \mathbb{R}^{d_{l}}, 
\end{gather} 
\newcommand{\target}{\text{target}}
\newcommand{\NS}{\text{NS}}
\newcommand{\PT}{\text{PT}}
\vspace{-6mm}
\subsection{Optimization of Adaptive-masking for Graph Embedding}
This subsection describes three loss functions used in our method.
In the training phase, we connect $\vhtarget$ in the target encoder stream to the LLM and optimize $\paramte$ with the prompt tuning loss $L_{\PT}$.
During training the target encoder with $L_{\PT}$, the concept encoder-decoder module is optimized exclusively with $L_{\target}$, in a JEPA approach.
Once entire network has been trained, we connect $\vhout$ to the downstream LLM rather than $\vhtarget$ at the inference phase.
One challenge of AGE lies in optimizing $\paramms$.
Optimizing $\paramms$ using $L_{\target}$ is difficult due to the non-differentiability of the sampling operation.
Therefore, we propose an additional loss function $L_{\NS}$ for optimizing $\paramms$.
\\
\textbf{Prompt tuning loss} $L_{\PT}$ maximizes accuracy of a downstream task. It was originally introduced in \cite{G-Retriever}, and we use the definition as is.
We optimize $\paramte$, $\paramgd$, and $\paramproj$ with $L_{\PT}$. 
The concrete implementation depends on the benchmark tasks; refer to the original papers for details. 
Note that we train $\paramge$ with $L_{\target}$ rather than $L_{\PT}$, as the concept encoder-decoder is used in the inference phase and the upstream network should be optimized to that module.
\\
\textbf{Target loss $L_{\target}$} optimizes parameters $\paramge$, $\paramce$, and $\paramcd$ to maximize embedding reconstruction by minimizing the distance between $\vhout$ and $\vhtarget$ for each auxiliary node indexed by ${I}_{\text{aux}}$ as follows:
\begin{gather} 
  L_{\target}(\paramge,\paramce,\paramcd) = \frac{1}{N_{\text{aux}}} \sum_{i \in I_{\text{aux}}} \parallel {h^i_{\text{out}}} - sg({h^i_{\text{target}}}) \parallel_{2},
\end{gather}
with $N_{\text{aux}}$ defined as $N - N_{\text{key}}$. The objective is to apply knowledge distillation to effectively represent key nodes for reconstructing auxiliary nodes. Furthermore, we apply normalization to enhance stability during the learning process, details are provided in the Appendix Table B.8.
\\
\textbf{Sampling loss $L_{\NS}$} optimizes $\paramms$ using RL-inspired supervision.
Regarding the operation as an action, the node sampler as a policy network, and $\vhout$ as a state, we design $L_{\NS}$ on $\prms=\{p^1,\ldots,p^N\}$ in Eq. \ref{eq:node_sampler} as follows:
\begin{gather}
  L_{\NS}(\paramms)=-\frac{1}{N_{\text{aux}}}\sum_{i\in I_{\text{aux}}}\left( \log(p^i) \times sg(\parallel h^i_{\text{out}}-h^i_{\text{target}}\parallel_{2})\right).
\end{gather}
The loss is back-propagated only to $\paramms$. Here, for each node assigned to $I_{\text{aux}}$, larger $\parallel h^i_{\text{out}}-h^i_{\text{target}}\parallel_{2}$ increases $p^i$ more, resulting in pushing such node into $I_{\text{key}}$.
This reflects our aim to classify nodes that are difficult to predict from their surrounding as {\it key nodes}. Additional strategies for key node selections and sampling optimization for RL are discussed in the Appendix Table B.6.
\newline
\textbf{Total loss} is given as $L_{\PT} + L_{\target} + L_{\NS}$. Since we designed the optimization process so that each loss optimizes different modules without overlap, our methods does not require weight adjustment between these losses.
%
%
%
%
%
%
%
%
%
\vspace{-3mm}
\subsection{Analysis of Learning Objectives}
To explain the motivation for architecture design and applying a distributed loss to each module, we analyze the learning objective of the $R{\omega}$ module, which represents expected $\overline{S^{*}}$ for LLM $\pi_{\theta}$ on LoRA finetuning as:
\begin{equation}
\begin{aligned}
\mathcal{L}
&= \underbrace{
-\mathbb{E}_{(S^{*})}\!\left[\log R{\omega}(\overline{S^{*}} \mid S^{*})  \right]
}_{\text{Loss of Graph Representation Module}}
\\
&
\underbrace{
-\mathbb{E}_{(q, T^{*},\overline{S^{*}})}\!\left[\log \pi_{\theta}(i \mid q, T^{*},\overline{S^{*}})\,  \pi_{\theta}(r \mid q, T^{*},\overline{S^{*}}, i)\right]
}_{\text{Loss of LLM}} 
\end{aligned}
\end{equation}
According to Bayes' Theorem, given an input $X$, a target $Y$, and latent rationales $Z$, we can sample these latent rationales $Z$ from the posterior distribution $P(Z|X, Y)$. This posterior represents the probability of latent $Z$ given both the input $X$ and the target $Y$. To compute the marginal likelihood of obtaining answer $Y$ given input $X$, we marginalize over all possible rationales $Z$:
\begin{equation}
\begin{aligned}
P(Y|X)
= \sum_{Z \sim P(Z|X,Y)} P(Z, Y|X)
=\sum_{Z \sim P(Z|X,Y)} P(Z|X) \cdot P(Y|X, Z)
\end{aligned}
\end{equation}
The equations above show how to compute the marginal likelihood $P(Y|X)$. Equation (15) with first line makes explicit that $Z$ is sampled from the posterior distribution $P(Z|X,Y)$. Second line extend applies the chain rule of probability to decompose $P(Z, Y|X)$ into two components: $P(Z|X)$ and $P(Y|X,Z)$.
Following this analysis, we apply it to the learning objective for target representation $\overline{S^{*}}$ given input $S^{\*}$, latent $\mathcal{Z}$ from a posterior $R{\omega}(\mathcal{Z}|S^{\*}, \overline{S^{\*}})$ that bridges $S^{\*}$ and $\overline{S^{\*}}$. The marginal likelihood of $\overline{S^{\*}}$ given $S^{\*}$ is:
\begin{equation}
\begin{aligned}
R\omega(\overline{S^*}|S^*)
&=\sum_{Z\sim R\omega(Z|S^*,\overline{S^*})}R\omega(Z,\overline{S^*}|S^*) 
\\
&
=\sum_{Z\sim R\omega(Z|S^*,\overline{S^*})}R\omega(Z|S^*)\cdot R\omega(\overline{S^*}|S^*,Z)
\end{aligned}
\label{eq:}
\end{equation}
Above analysis shows that learning objective Graph Representation implicitly learns to identify the latent $Z$ and map it to the expected $\overline{S^{*}}$ for LLM:
\begin{equation}
\begin{aligned}
-\mathbb{E}\!\left[\log_{R{\omega}}(\overline{S^{*}} \mid S^{*})  \right] 
&=
\underbrace{
\mathbb{E}
\!\left[ \log R{\omega}(\mathcal{Z} \mid S^{*},\overline{S^{*}}) \right] 
}_{\text{Loss of Latent Identification}}
\; \;
\\
&
\underbrace{
-\mathbb{E}\!\left[
\log R{\omega}(\mathcal{Z} \mid S^{*}) \cdot R{\omega}(\overline{S^{*}} \mid S^{*},\mathcal{Z}))
\right]
}_{\text{Loss of Representation}}\\
\end{aligned}
\label{eq:}
\end{equation}
Instead of using a single model for both latent $\mathcal{Z}$ identification and $\overline{S^{*}}$ representation learning. We separate the learning into $Sampler_{\theta}$ for \textbf{latent identification by minimizing reconstruction loss} and $Encoder_{\theta}$-$Decoder_{\theta}$ for representation as:
\begin{equation}
\begin{aligned}
-\mathbb{E}\!\left[\log_{R{\omega}}(\overline{S^{*}} \mid S^{*})  \right] 
&\approx
\underbrace{
-\mathbb{E}_{(S^{*},\overline{S^{*}})}\!\left[ V_{key} \in \mathcal{Z} \sim \log\ Sampler_\theta (\mathcal{Z} \mid S^{*},\overline{S^{*}}) \right] 
}_{\text{Loss of Node Sampling}}
\\
&\underbrace{
-\mathbb{E}_{(S^{*})}
\!\left[
\log Encoder_{\theta} (\mathcal{Z} \mid V_{key}) \cdot Decoder_{\theta} (\overline{S^{*}} \mid \Delta_{V_{masked}},\mathcal{Z}) 
\right]
}_{\text{Loss of Encoder-Decoder}}
\end{aligned}
\label{eq:}
\end{equation}
{By separating the learning processes, the target encoder learns the representation directly, while the encoder-decoder learns to reconstruct this representation through
\textbf{Evidence Lower Bound (ELBO) optimization}. 
Specifically, the node sampler learns to extrapolate $V_{key} \in \mathcal{Z}$, making static sampling illogical. Therefore, our node sampler with encoder-decoder architecture and explicit loss distribution yields efficient learning signals, faster convergence, and improved graph representations.
To support our analysis, we provide empirical comparisons of sampling strategies in Appendix Figure B.3, B.4 and analyze the stability of the target encoder teacher module for the encoder-decoder in Appendix Table B.8. In the prompt tuning setting, given $r$ as the reasoning trajectory, the learning objective for frozen LLMs with graph representation model $R_{\omega}$ is:
\begin{equation}
\begin{aligned}
\mathcal{L}
&=
\underbrace{
\mathbb{E}
\!\left[ \log R_{\omega}(\mathcal{Z} \mid S^{*},\overline{S^{*}}) \right] 
}_{\text{Loss of Latent Identification}}
\; \;
\underbrace{
-\mathbb{E}\!\left[
\log R_{\omega}(\mathcal{Z} \mid S^{*}) \cdot R_{\omega}(\overline{S^{*}} \mid S^{*},\mathcal{Z}))
\right]
}_{\text{Loss of Representation}}\\
&
\underbrace{
\underbrace{
-\mathbb{E}\!\left[\log \pi_{\theta}(i \mid q, T^{*},\overline{S^{*}})\right]
}_{\text{Loss of Knowledge Recalling}}
\; \;
\underbrace{
-\mathbb{E}\!\left[\log \pi_{\theta}(r \mid q, T^{*},\overline{S^{*}}, i)\right]
}_{\text{Loss of Contextualized Reasoning}}
}_{\text{Frozen}} 
\end{aligned}
\end{equation}
{We observe that $R_\omega$ explicitly learns to identify the latent $\mathcal{Z}$ from $S^*$ for LLM-expected $\overline{S^*}$. 
During training with frozen LLM parameters, $R_\omega$ implicitly captures latent identification $i$ by satisfying LLM's expectations: $Retriever(S^* \mid q, T^*) \; R_\omega(\overline{S^*} \mid S^*, \mathcal{Z}) \approx \pi_\theta(\overline{S^*} \mid q, T^*, i)$, yielding $\mathcal{Z} \subseteq i$. Therefore, $R_\omega$ able to learns a subspace of the frozen LLM's complete latent space through this objective. 
Throughout this, we argue that leveraging a learned latent space $\mathcal{Z}$, robustly restructured into the LLM-expected representation $\overline{S^*}$, can directly improve knowledge recall and indirectly enhance reasoning.
%
%
%
%
%
%
%
%
%
%
\vspace{-3mm}
\section{Experiments}\label{Experiments1}
\vspace{-3mm}
%
\textbf{Datasets and Evaluation Metrics}. 
Following previous work \cite{G-Retriever,grag_graph_retrieval_augmented_generation,ji2024ntllm}, we conduct experiments on ExplaGraphs~\cite{saha2021explagraphs} is a generative commonsense reasoning dataset, SceneGraphs~\cite{hudson2019gqa} is a visual question answering dataset. And  WebQSP~\cite{yih2016value},  ComplexWebQuestions (CWQ) \cite{talmor2018web} is a large question-answering dataset derived from Web questions, where all queries can be answered using Freebase, a large collaborative knowledge graph database. We use accuracy as the primary metric for ExplaGraphs and SceneGraphs, datasets focusing on reasoning, following \cite{G-Retriever,grag_graph_retrieval_augmented_generation,ji2024ntllm}.
For WebQSP and CWQ, a dataset with extra-large graphs, we use the Hit@1 metric, as in \cite{reasoning_on_graphs}. 
%
%
%
\newline
\textbf{Implementation Details\label{setting1}}.
We employed the open-source Llama3.2 (1B and 3B) \cite{llama3.2}, Llama 2 (7b and 13B) \cite{touvron2023llama} and Llama3.1 (8B) \cite{llama3.1} as frozen LLM components. Based on the analysis in \ref{ablationstudy}, we set the sampling rate $\rho=0.3$ (see Appendix 2.1 and 2.5 for cost–benefit trade-offs).
\vspace{-3mm}
\subsection{Main Results}\label{mainresults}
Table \ref{tab:performance} illustrates our main results, comparing the methods in three settings: 
\textbf{Frozen LLM with Graph Embedding}: Use a graph embedding technique to tune tokens with given prompt. \textbf{Frozen LLM with Graph Embedding + PEFT}: Apply LoRA \cite{hu2021lora}, a PEFT technique, in combination with the graph embedding technique. \textbf{LLM with LLM-Retriever}: Use LLM for retrieval in addition to the one for inference. Any methods train either LLM. These are reference scores with a larger computational cost.
%
\begin{table}[tb]
    \centering
    \caption{Performance comparison across ExplaGraphs, SceneGraphs, and WebQSP datasets under the five settings. The best and second-best scores are highlighted in \textbf{bold} and \underline{underline}, respectively.}
    \small
    \begin{tabular}{lllcccc}
        \toprule
        
        Setting & Method  & LLM & \makecell{Expla\\ Graphs} & \makecell{Scene\\ Graphs} & WebQSP & CWQ\\
        
        \midrule
        \multirow{3}{*}{\makecell[l]{Frozen LLM \\ w/ Graph \\ Embedding}} 
         
        & G-Retriever & Llama3.2-1B  & $0.5595$ & $0.7540$ & $60.1 $ & $-$ \\
        & G-Retriever & Llama3.2-3B  & $0.7761$ & $0.8229$ & $71.3 $ & $-$ \\
        & G-Retriever & Llama2-7B    & $0.8516$ & $0.8131$ & $68.1 $ & $-$ \\
        
        & \textbf{AGE G-Retriever}  & Llama3.2-1B  &  0.8267 & 0.8184 & 62.5 & $-$ \\
        & \textbf{AGE G-Retriever}  & Llama3.2-3B  &  0.9260 & 0.8930 & 73.5 & $-$ \\
        & \textbf{AGE G-Retriever}  & Llama3.1-8B  &  \underline{0.9350} & 0.9276 & \underline{78.3} & $-$ \\
        
        

        \midrule
        \multirow{3}{*}{\makecell[l]{Frozen LLM \\ w/ Graph \\ Embedding \\ \hspace{0.5em} + PEFT(LoRA)}}
        
        & G-Retriever & Llama3.2-1B & $0.7328$ & $0.8689$ & $65.3$ & $-$ \\
        & G-Retriever & Llama3.2-3B & $0.8339$ & $0.9074$ & $71.4$ & $-$ \\
        & G-Retriever & Llama2-7B   & $0.8705$ & $0.8683$ & $70.2$ & $-$ \\
        

        & \textbf{AGE G-Retriever}  & Llama3.2-1B & $0.8501$ & $0.9056$ & $69.1$ & $-$ \\
        & \textbf{AGE G-Retriever}  & Llama3.2-3B & $0.9134$ & \textbf{0.9486} & $77.3$ & $-$ \\
        & \textbf{AGE G-Retriever}  & Llama3.1-8B & \textbf{0.9612} & \underline{0.9325} & 80.3 & $-$ \\

        & AMAR  & Llama2-7B   & $- $ & $- $ & 84.3    & 82.9 \\
        & AMAR  & Llama2-13B  & $- $ & $- $ & 83.3    & 83.1 \\

        & \textbf{AGE AMAR}  & Llama2-7B   & $- $ & $- $ & 86.5    & \textbf{85.2} \\
        & \textbf{AGE AMAR}  & Llama2-13B  & $- $ & $- $ & 86.2    & \underline{85.1} \\

        \midrule
        \multirow{3}{*}{\makecell[l]{LLM\\ w/ LLM Retriever}}
        & ToG              & GPT-4          & $- $ & $- $ & 82.6    & 67.6 \\ 
        & ReKnoS           & GPT-4          & $- $ & $- $ & 84.9    & 68.2 \\
        & KG-Agent         & Llama2-7B      & $- $ & $- $ & 83.3    & 72.2 \\
        & Paths-over-Graph & GPT-4          & $- $ & $- $ & \textbf{96.7}    & 81.4 \\
        & Plan-on-Graph    & GPT-4          & $- $ & $- $ & 87.3    & 75.0 \\
        & DoG              & GPT-4          & $- $ & $- $ & \underline{91.0}    & 56.0 \\
        \bottomrule
    \end{tabular}
    \label{tab:performance}
\end{table}
%
%
Among \textbf{Frozen LLM with Graph Embedding} settings (with and without PEFT), \textbf{AGE consistently improved performance of G-Retriever and AMAR regardless of the backbone LLM models}.
Without PEFT, Llama3.2-1B with AGE showed the most notable gain against G-Retiever: $26.72$ percent points increase on ExplaGraphs, while the least gain was observed with Llama3.2-3B on WebQSP, which was $2.02$ points. This might be due to the extra-large size of knowledge graphs in WebQSP datasets, which include textual knowledge absent at pretraining, and non-parameter retriever struggling to provide critical information for representation. AGE maintains consistent superiority against G-Retriever and shows more gains on retrieval from smaller graphs.
By employing a cross-question approach enriched with retrieved elements, GRAG improved performance by $2.8$ points, AMAR achieved an improvement of $4.2$ points with its baseline. This suggests that improving embedding module is beneficial, it alone may not be enough to boost performance significantly, and relying on a non-parameter retriever could be limiting. Despite that challenge, when integrated with AMAR, AGE continues to achieve further enhancements the performance.
%
Direct comparison with current \textbf{LLM with LLM Retriever} methods on WebQSP and the larger CWQ shows that AGE AMAR, which applies non-parametric retriever-based approaches, outperforms the proprietary LLM-based retriever ReKnoS \cite{wang2025reasoning} on both datasets. AGE AMAR underperforms compared with Paths-over-Graph \cite{tan2025paths}, Plan-on-Graph \cite{wu2024planongraph} and DoG \cite{ma2025debate} on WebQSP, but outperforms them on CWQ,  suggesting that AGE AMAR is advantageous on larger datasets, showing substantial potential for future work (as reported in the Appendix Table B.11, further including the more baseline methods.) 
%
%

\subsection{Ablation study}\label{ablationstudy}
\begin{table*}[tb]
\centering
\small
\caption{Performance improvements (Llama3.2 1B, ExplaGraphs; \% points).}
\begin{tabular}{c|c|c|c|c|c}
\toprule
      & G-Retriever & \makecell{GA w\\ Random mask} & \makecell{JEPA w\\ Random mask} & \makecell{GA w\\ Node sampler} & \makecell{JEPA w\\ Node sampler} \\
\midrule
Loss  & $L_{\text{PT}}$
      & $L_{\text{PT}} + L_{\text{target}}$
      & $L_{\text{PT}} + L_{\text{target}}$
      & \makecell{$L_{\text{PT}} + L_{\text{target}}$ \\ $+\, L_{\text{NS}}$}
      & \makecell{$L_{\text{PT}} + L_{\text{target}}$ \\ $+\, L_{\text{NS}}$} \\
\midrule
Acc   & 0.5595
      & \makecell{0.6532 \\ ($\uparrow$ 9.37\%)}
      & \makecell{0.7141 \\ ($\uparrow$ 15.46\%)}
      & \makecell{0.7870 \\ ($\uparrow$ 22.75\%)}
      & \makecell{0.8267 \\ ($\uparrow$ 26.72\%)} \\
\bottomrule
\end{tabular}
\label{table:Architecture}
\end{table*}
\textbf{Performance Comparison of Self-Supervised Learning Architectures.}
Table \ref{table:Architecture} presents the performance of concept encoder-decoder with some variations and the baseline of G-Retriever, which demonstrates contribution of proposed technique independently. As a SSL variation, we prepared AGE based on generative architecture (GA) against our choice of JEPA. We also compared AGE with random mask against the proposed learnable node sampler. All the results used Llama3.2 1B as its LLM. Using a GA with a random mask, AGE achieves a performance of $0.6532$, which is a $9.37\%$ improvement over the baseline. Next, AGE with a random mask, improving performance by $15.46\%$ over the baseline. Finally, AGE using JEPA with the learnable node sampler achieves a $26.72\%$ improvement over the baseline. Based on these experiments, we confirmed that JEPA works better than GA as expected, while node sampler further improves performance with a notable margin.
Furthermore, we include an additional study on architectural design choices in the Appendix Table B.7, such as reason we choose different input features for LLMs during training and inference in GraphRAG.
\begin{figure*}[t]
\centering
\small
\begin{minipage}{0.49\textwidth}
\centering
\begin{tikzpicture}
\begin{axis}[
xlabel={Sampling rate $\rho$},
ylabel={Accuracy},
xtick={0.3, 0.4, 0.5, 0.6, 0.7},
legend pos=south west,
ymin=0.7, ymax=0.95,
grid=major,
width=\textwidth,
height=4.5cm
]
\addplot[
color=blue,
mark=square,
]
coordinates {
(0.7, 0.78339)
(0.6, 0.79422)
(0.5, 0.80685)
(0.4, 0.80505)
(0.3, 0.81407)
(0.25, 0.81046)
};
\addlegendentry{Llama3.2 1b}
\addplot[
color=red,
mark=triangle,
]
coordinates {
(0.7, 0.91516)
(0.6, 0.87545)
(0.5, 0.91516)
(0.4, 0.90613)
(0.3, 0.92599)
(0.25, 0.88447)
};
\addlegendentry{Llama3.2 3b}
\end{axis}
\end{tikzpicture}
\caption{Performance of against sampling rate (ExplaGraphs)}
\label{fig:explain_graph}
\end{minipage}
\hfill
\begin{minipage}{0.49\textwidth}
\centering
\begin{tikzpicture}
\begin{axis}[
xlabel={Sampling rate $\rho$},
ylabel={Hit@1},
xtick={0.3, 0.4, 0.5, 0.6, 0.7},
legend pos=south east,
ymin=50, ymax=75,
grid=major,
width=\textwidth,
height=4.5cm
]
\addplot[
color=blue,
mark=x,
]
coordinates {
(0.7, 61.5479)
(0.65, 62.1622)
(0.6, 60.1351)
(0.55, 60.1351)
(0.5, 60.6880)
(0.45, 62.0393)
(0.4, 58.5381)
(0.35, 60.9337)
(0.3, 62.5307)
(0.25, 62.3464)
};
\addlegendentry{Llama3.2 1b}
\addplot[
color=red,
mark=o,
]
coordinates {
(0.7, 71.8673)
(0.65, 68.6732)
(0.6, 71.2359)
(0.55, 71.9902)
(0.5, 69.457)
(0.45, 71.437)
(0.4, 71.498)
(0.35, 73.464)
(0.3, 72.213)
(0.25, 71.482)
};
\addlegendentry{Llama3.2 3b}
\end{axis}
\end{tikzpicture}
\vspace{-3mm}
\caption{Performance against sampling rate (WebQSP)}
\label{fig:webqsp}
\end{minipage}
\vspace{-3mm}
\end{figure*}
\begin{figure*}[t]
\begin{center}
\includegraphics[width=0.95\linewidth ]{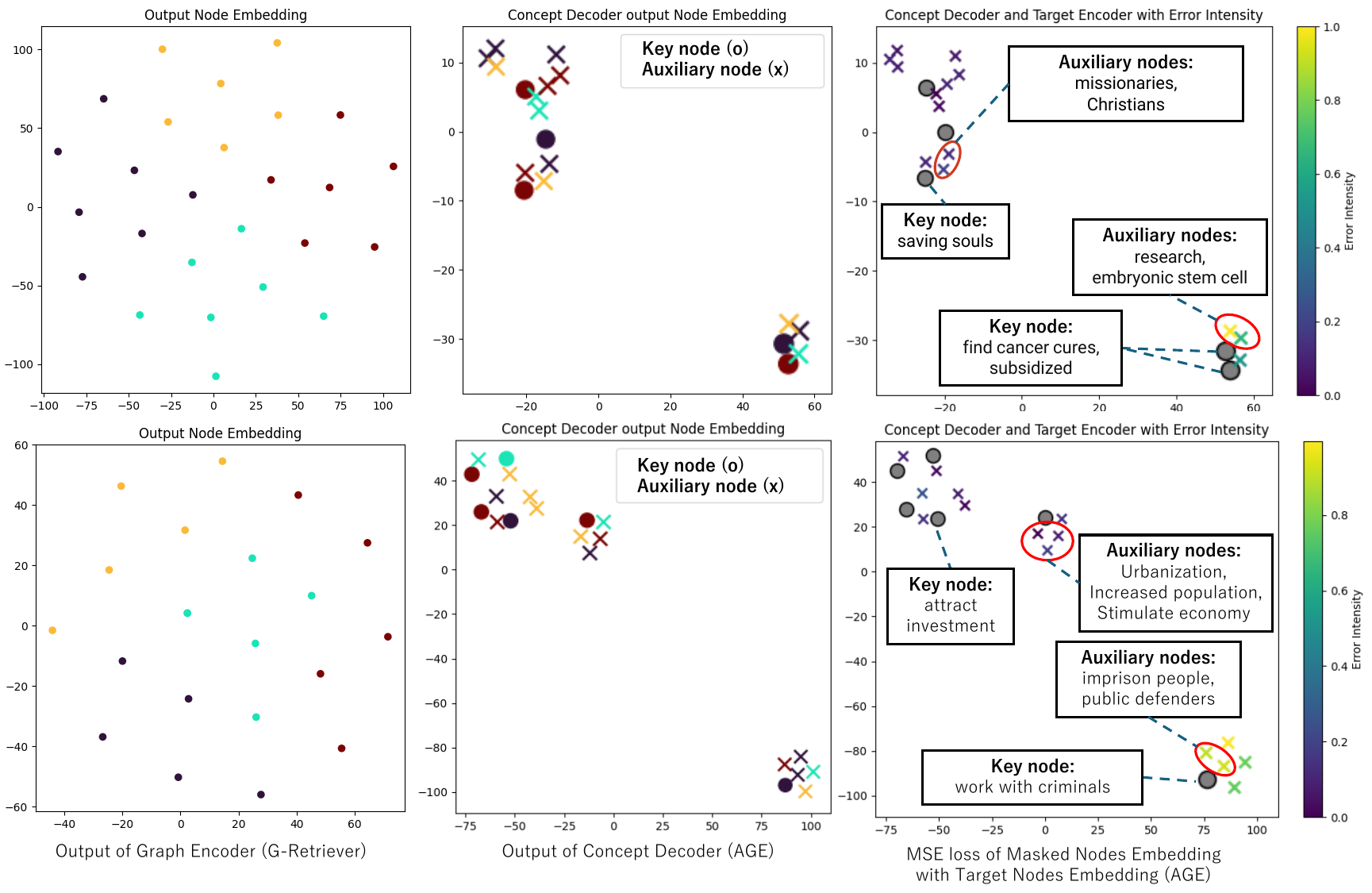}
\end{center}
\vspace{-3mm}
   \caption{Node embedding of G-Retriever and AGE G-Retriever with sampling rate $\rho=0.3$ using t-SNE \cite{vanDerMaaten2008}: Nodes are colored by clustering node's text (left two), or target error (the rightest).}
\label{fig:Qualitative0}
\vspace{-3mm}
\end{figure*}
\newline
\textbf{Analysis on the Sampling Rate $\rho$}.
Figure \ref{fig:explain_graph} illustrates how the sampling rate impacts AGE G-Retriever performance on ExplaGraphs and WebQSP, guiding our hyper-parameter setting. Average retrieved nodes are $18.21$ and $5.17$ on WebQSP and ExplaGraphs, respectively.
A sampling rate $\rho=0.3$ gives the best performance on ExplaGraphs with both LLM settings ($81.4\%$ for Llma3.2-1B and $92.6\%$ for Llama3.2-3B).
The same setting also achieves the best performance on WebQSP for Llama3.2-1B ($62.5\%$) and the second best for Llam3.2-3B ($72.2\%)$, compared to the best score of $73.5\%$ at $\rho=0.35$.
From these observations, we decided to use $\rho=0.3$ through the experiments.
\newline
\textbf{Other LLM backbones.}
A Qwen3.5 result (Table~\ref{tab:qwen}) confirms the same trend. We will clarify the LoRA setting. AGE consistently outperforms G-Retriever across all scales and tasks, with especially large gains on ExplaGraphs. The improvements remain strong even at the smaller 0.8B scale, indicating efficiency and scalability.
Gains on WebQSP further demonstrate the robustness of the proposed method across different tasks.
\begin{table*}[t]
{
\centering
\renewcommand{\thetable}{R1}
\caption{The performance on Qwen3.5 family.}
\label{tab:qwen}
\setlength{\aboverulesep}{0pt}
\setlength{\belowrulesep}{1pt}
\renewcommand{\arraystretch}{0.85}
\small
\setlength{\tabcolsep}{4.5pt}
\begin{tabular}{c r cc cc}
\toprule
 &  & \multicolumn{2}{c}{\textbf{Frozen}} & \multicolumn{2}{c}{\textbf{LoRA}} \\
\cmidrule(lr){3-4} \cmidrule(lr){5-6}
\textbf{Method} & \multicolumn{1}{c}{\textbf{Size}} & \textbf{Expla Graphs} & \textbf{WebQSP} & \textbf{Expla Graphs} & \textbf{WebQSP} \\
\midrule
\multirow{2}{*}{G-Retriever}
    & 0.8B & 0.4832 & 59.5 & 0.7178 & 65.0 \\
    & 2B   & 0.7477 & 72.8 & 0.8298 & 69.9 \\
\midrule
\multirow{2}{*}{\textbf{AGE (ours)}}
    & \textbf{0.8B} & \textbf{0.8089} & \textbf{61.7} & \textbf{0.8231} & \textbf{68.4} \\
    & \textbf{2B}   & \textbf{0.9101} & \textbf{73.6} & \textbf{0.9097} & \textbf{77.3} \\
\bottomrule
\end{tabular}
}
\end{table*}
\newline
\textbf{Weighting Loss.}
The three losses optimize disjoint parameters, so weighting is unnecessary by design; confirmed also in Table~\ref{tab:1}.
Equal weighting (1:1:1) consistently yields the best performance across datasets and model sizes.
Deviating from equal weights degrades performance, with no consistent benefit from emphasizing any single loss.
This supports that the three losses operate on disjoint parameters, making explicit weighting unnecessary.
\begin{table}[t]
\centering
{
\renewcommand{\thetable}{R2}
\caption{Loss-weight ablation ($w_i$ on $L_{NS}, L_{target}, L_{PT}$).}
\setlength{\aboverulesep}{0pt}
\setlength{\belowrulesep}{1pt}
\renewcommand{\arraystretch}{0.85}
\small
\setlength{\tabcolsep}{9pt}
\begin{tabular}{c cc cc}
\toprule
\multirow{2}{*}{\textbf{$w_1 : w_2 : w_3$}} & \multicolumn{2}{c}{\textbf{ExplaGraphs}} & \multicolumn{2}{c}{\textbf{WebQSP}} \\
\cmidrule(lr){2-3} \cmidrule(lr){4-5}
 & \textbf{1B} & \textbf{3B} & \textbf{1B} & \textbf{3B} \\
\midrule
1.3 : 0.7 : 1 & 0.8087 & 0.8863 & 59.4 & 72.0 \\
0.7 : 1.3 : 1 & 0.7834 & 0.8971 & 59.6 & 72.4 \\
1 : 1 : 1     & \textbf{0.8267} & \textbf{0.9260} & \textbf{62.5} & \textbf{73.5} \\
\bottomrule
\end{tabular}
\label{tab:1}
}
\end{table}
\newline
\textbf{Qualitative Evaluation.\label{QE0}} 
To analyze node sampling results, we visualized the node sampling results on two samples from the ExplaGraphs test set in Figure \ref{fig:Qualitative0}.
Nodes are colored based on clustered text embeddings to track node-wise feature restructuring through graph relationships.
The graph encoder process maintains the clustering structure of text graph embeddings in both G-Retriever (first column) and AGE (second column).
In contrast, the concept encoder-decoder module (second column) shuffles the colored nodes, indicating a reorganization of the node-wise embeddings.
\newline
The last column displays text entities of some key and auxiliary nodes. 
Our node sampler is designed to sample entities from specific domains as key nodes.
As "saving souls" is sampled as a key node, inferring "missionaries" and "Christians" from it seems easier than the reverse. We observe the same tendency with the key node "work with criminals" and the auxiliary nodes "imprison people" and "public defenders."
The last column also shows the target loss for each auxiliary node using the color bar, where 1.0 represents the maximum error in the test set, and 0.0 the least. From the color visualization, we observe that non-isolated key nodes achieve lower errors in auxiliary node prediction, suggesting that relations between key nodes support the prediction. We provides additional qualitative results, including failure cases, in Appendix B.9.
%
\vspace{-3mm}
\section{Limitation and Conclusion} \label{Limit} 
\vspace{-3mm}
Although our method delivers substantial improvements in GraphRAG, several limitations remain.
First, we used a fixed sampling rate despite variations in key node density between the graphs. 
Second, we tested AGE only on GraphRAG tasks, even though it is applicable to other modalities. These limitations suggest areas for future improvement and the potential for broader applications.
Third, our approach primarily targets small-scale models, and its effectiveness for large-scale LLMs remains unexplored due to computational constraints. Finally, our method focuses on representing retrieved structured data for LLMs rather than directly addressing graph learning tasks (e.g., node classification or link prediction). In the absence of theoretical guarantees on the benefits of node and link integration, our current scope is mainly limited to KGQA scenarios.
\\
We proposed Adaptive-masking for Graph Embedding (AGE) to improve structured graph embeddings and enhance LLM performance on GraphQA tasks.
The method introduced JEPA, a self-supervised learning architecture which enhanced the graph-structure embedding for downstream reasoning tasks.
Our node sampler demonstrated its effectiveness in the ablation study, successfully identified key nodes within given graphs.
The quantitative results confirmed AGE's consistent performance gain in GraphRAG tasks while maintaining computational cost.
We hope this work contributes to structured knowledge representation for intelligent agents and facilitates cross-modal reasoning through structured perceptual representations.



%
%
\bibliographystyle{splncs04}
\bibliography{main}

\newpage
\clearpage



\title{Appendix \\ AGE: Adaptive-masking for Graph Embedding in Graph Retrieval-Augmented Generation}

\author{Bao Long Nguyen Huu \inst{1} \and
Atsushi Hashimoto\inst{2}}

\authorrunning{F.~Author et al.}

\institute{OMRON Corporation \and
OMRON SINIC X Corporation}

\maketitle

\renewcommand{\thefigure}{\Alph{section}.\arabic{figure}}
\renewcommand{\thetable}{\Alph{section}.\arabic{table}}

\section{Proof of Concept} 
In this section, we motivate AGE's design for representing retrieved subgraphs for higher-order reasoning skills. We assume this requires on the most common static search single-turn retrieval and LLM for academic simplicity. 
\\
\textbf{Problem definition:} Given input query $q$, we generate chain-of-thoughts response $y = i \oplus r$ (interleaved domain knowledge $i$ and reasoning steps $r$). We use a static retrieval engine that returns $T^{*}$ for text knowledge and $S^{*}$ for subgraph. The learning objective $\min_{\theta} \mathbb{E}[\mathcal{L}]$ optimizes representation $S^{*}$ to prioritize reasoning $r$ over knowledge $i$.
\\
For the chain-of-thoughts response is defined as $y = i \oplus r$ where $y$ is the concatenation of knowledge $i$ and reasoning $r$ through three discrete generation processes below.
\begin{itemize}
\item \textbf{Graph Knowledge Retrieval:}
Given the query $q$ on a textual graph $G$ and $S(G)$ is the set of all subgraphs of $G$. Retrieval system extracts relevant subgraph $S^{*} \in S(G)$ and text-modal knowledge $T^{*}$. Popular retrieval systems select top-k elements by cosine similarity, yielding nodes $V_k^{*}$ and edges $E_k^{*}$ considered relevant to the query. A non-optimized retriever may yield corrupted subgraphs $S^{*}=(V_k^{*},E_k^{*})$, as they may be contain redundant or lack suggestive elements.

\item \textbf{Graph Knowledge Representation:}
Graph embedding module is trained to represent graph that guides the LLM to produce expected answers. Embedding module $R_\omega(\cdot)$ learn to represent corrupted subgraph $S^{*}$ to $\overline{S^{*}}$ for an LLM$\Phi$ improve generation.

\item \textbf{Contextualized Reasoning:}
Given $q$, $T^{*}$ and $\overline{S^{*}}$, LLM synthesizes domain knowledge $t$ by recall their internal parametric knowledge with external inputs, following the conditional distribution $i \sim \pi_{\theta}(i|q, T^{*},\overline{S^{*}})$. Then LLM generates reasoning steps $r$ conditioned on $q$, $T^{*}$, $\overline{S^{*}}$ with the recalled internal knowledge $i$, adhering to the reasoning distribution $r \sim \pi_{\theta}(r|q, T^{*},\overline{S^{*}}, i)$.

\end{itemize}
Here we formally analyze and discuss the subgraph representation learning objectives of both vanilla embedding module and AGE embedding module with LLM generation distribution.

\begin{itemize}

\item Subgraph embedding module that employ GNN or Transformer:
\begin{align}
\overline{S^{*}} &= \text{R}_\omega^{\text{GNN}}(S^{*}) = \big\{\, \overline{V}_i \,\big\}_{i \in V^{*}}
\\
\overline{V}_i
&= \sum_{j \in \mathcal{N}(i)} \alpha_{ij}\, W_V V_j,
\end{align}
Here, $\mathcal{N}(i)$ denotes the local neighborhood of node $i$ (e.g., $\mathcal{N}(i)=\{j \mid (i,j)\in E^{*}\}$).
\begin{align}
\overline{S^{*}}
&= \text{R}_{\omega}^{\mathrm{Transformer}}(S^{*})
 = \big\{\, \overline{V}_i \,\big\}_{i \in V^{*}}
\\
\overline{V}_i
&= \sum_{j \in V^{*}} \alpha_{ij}\, W_V V_j,
\qquad
\alpha_{ij}
= \softmax_{j \in V^{*}}
\!\Big( \tfrac{(W_Q V_i)^\top (W_K V_j)}{\sqrt{d_k}} \Big),
\end{align}
where $W_Q,W_K,W_V$ are learnable matrices, $d_k$ is the key dimension.
These formulations use attention-weighted aggregation as node embeddings. Static retrieval from graphs with high-order structural patterns (motifs/role patterns) produces corrupted subgraphs that contain redundant nodes or miss critical elements. 
Without explicit structural constraints, training weighted-sum aggregation through supervised or semi-supervised learning separates signal from noise, may lead to diluted node representations within the embedding space.
In a frozen state, LLMs may fail to capture relationships in the embedding space due to their struggle to handle diluted node representations.

\item Subgraph embedding module that employ RL-guide mask-based SSL:
\begin{align}
\overline{S^{*}} 
= R{\omega}^{\mathrm{AGE}}(S^{*}) 
= \big\{\, \overline{V}_i \,\big\}_{i \in V^{*}}
=Decoder_{\phi}\big(\Delta_{V_{masked}}, Encoder_{\theta}(V_{key})\big) \\
= V_{key} \sim \text{Sampler}_\theta(V_{key} \mid V^{*})  \!\Bigg[  \sum_{j \in V^{*}} \alpha_{ij}  W_V \bigg[ \Big[\sum_{j \in V^{key}} \alpha_{ij}\, W_V V_j \Big]; \Delta_{V_{masked}}   \bigg] \!\Bigg]
\end{align}
$\Delta_{V_{masked}}$ denotes the auxiliary masked node features used as decoder input, while $V_{key}$ denotes the key node features used as encoder input and $[;]$ is the concatenation operation. 
\begin{align}
\begin{cases}
    \Delta_{V_{masked}} = Mask(V_{j})          & \text{if } a_{j} = 1 \\
    V_{key} = Visible(V_{j})       & \text{if } a_{j} = 0
\end{cases} \quad \sum_{j \in V^{*}} \text{Sampler}_\theta(a_{j} \mid V_j) ,
\end{align}
where $V_{key}$ represents key node features for encoder input, where $\text{Mask}(\cdot)$ masks features, $\text{Visible}(\cdot)$ preserves them, and $a_j \in \{0,1\}$ denotes the binary RL action ($1$=mask, $0$=keep) for node $j$.

By employing mask-based SSL alongside a reinforcement learning framework, AGE learns structural dependency node representations in the embedding space through reconstruction objectives. The reinforcement learning framework is used to estimate which nodes are critical for preserving graph structure and semantic information. Then, the estimated nodes are applied to guide mask-based SSL to reconstruct that provide structural constraints in the embedding space, enabling LLMs to better separate signals and capture relationships within it.

\item The joint generation distribution of LLMs is:
\begin{equation}
\begin{aligned}
\pi_{\theta}(y \mid q, T^{*},\overline{S^{*}})
&= \pi_{\theta}(i \oplus r \mid q, T^{*},\overline{S^{*}}) \\
&= \underbrace{\pi_{\theta}(i \mid q, T^{*},\overline{S^{*}})}_{\text{Knowledge Recalling}}
\cdot
\underbrace{\pi_{\theta}(r \mid q, T^{*},\overline{S^{*}}, i)}_{\text{Contextualized Reasoning}} ,
\end{aligned}
\label{eq:joint_generation}
\end{equation}

\item The loss function optimizes both knowledge integration and contextualized reasoning:
\begin{equation}
\begin{aligned}
\mathcal{L}
&= -\mathbb{E}_{(q, T^{*},\overline{S^{*}})}\!\left[\log \pi_{\theta}(i \mid q, T^{*},\overline{S^{*}})\, \pi_{\theta}(r \mid q, T^{*},\overline{S^{*}}, i)\right] \\
&= 
-\mathbb{E}_{(q, T^{*},\overline{S^{*}})}\!\left[\log \pi_{\theta}(i \mid q, T^{*},\overline{S^{*}})\right]
\; \;
-\mathbb{E}_{(q, T^{*},\overline{S^{*}})}\!\left[\log \pi_{\theta}(r \mid q, T^{*},\overline{S^{*}}, i)\right]
\, ,
\end{aligned}
\label{eq:loss_generation}
\end{equation}
\end{itemize}
Through the lens of multi-task learning, we compare above equations and from two perspectives:
\begin{itemize}
\item \textbf{Retrieved Subgraph Representation on LLMs' generation distribution.} Based on equation~\eqref{eq:joint_generation}, LLMs decompose response generation into knowledge recall and contextualized reasoning. In the frozen state, diluted subgraph representations trigger a cascade—weak knowledge recall causes flawed reasoning. This limitation creates a bottleneck that degrades response quality.

\item \textbf{Retrieved Subgraph Representation with Parameter-Efficient Fine-Tuning.}

Following previous research \cite{wang2025rare}, we assume the retrieved representation is \textbf{explicitly}, we have ~\eqref{eq:loss_generation} as:
\begin{equation}
\begin{aligned}
\mathcal{L}
&= \underbrace{-\mathbb{E}_{(q, T^{*},\overline{S^{*}})}\!\left[\log \pi_{\theta}(i \mid q, T^{*},\overline{S^{*}})\right]}_{\textbf{Loss of Integration}}\downarrow
\; \;
\underbrace{-\mathbb{E}_{(q, T^{*},\overline{S^{*}})}\!\left[\log \pi_{\theta}(r \mid q, T^{*},\overline{S^{*}}, i)\right]}_{\text{Loss of Reasoning}}\uparrow \, ,
\end{aligned}
\end{equation}
The arrows indicate the loss function shifts to reasoning. That means the loss term shifts from knowledge identification to integration, that $\pi_{\theta}(i \mid q, T^{*},\overline{S^{*}})$ has already reached "application" levels of retrieved graph knowledge. Therefore, explicit subgraph representation aids knowledge integration beyond mere identification during fine-tuning.

Conversely, the retrieved subgraph representation is \textbf{diluted}, we have ~\eqref{eq:loss_generation} as:
\begin{equation}
\begin{aligned}
\mathcal{L}
&= \underbrace{-\mathbb{E}_{(q, T^{*},\overline{S^{*}})}\!\left[\log \pi_{\theta}(i \mid q, T^{*},\overline{S^{*}})\right]}_{\textbf{Loss of Identification}}\uparrow
\; \;
\underbrace{-\mathbb{E}_{(q, T^{*},\overline{S^{*}})}\!\left[\log \pi_{\theta}(r \mid q, T^{*},\overline{S^{*}}, i)\right]}_{\text{Loss of Reasoning}}\downarrow \, ,
\end{aligned}
\end{equation}
The arrows indicate that the loss function prioritizes identification. This reveals inefficient knowledge use: the model identifies patterns directly, treating retrieved subgraphs as training data. This creates a trade-off: instead of learning to apply retrieved subgraphs in reasoning, $\pi_{\theta}(i \mid q, T^{*},\overline{S^{*}})$ prioritizes identifying graph structures over reasoning. Poor retrieved subgraph representation implicitly hinders reasoning capability development, forcing resources into identification tasks that better representations would cover. 
\end{itemize}

\section{Additional Experimental Details \label{Experiments2}} 

\subsection{Implementation Settings (AGE G-Retriever)}\label{ImpSet}
When integrated with G-Retriever, we consistently use the AdamW \cite{AdamW} optimizer and set the initial learning rate at $1e-4$, with a weight decay of $0.05$. Following the baseline work \cite{G-Retriever}, we set learning rate decays with a half-cycle cosine decay after the warm-up period. To avoid overfitting, we implement early stopping with a patience of $3$ epochs. The experiments used 2 NVIDIA 2080Ti-11G or 2 NVIDIA A100-80G GPUs.
\\
\textbf{GNN.} We use Graph Transformer as the GNN backbone applied in the Graph Encoder and Graph Structure Based Aggregator. Similar to previous approaches \cite{G-Retriever}, our settings for its employ $2$ layers, each with $4$ attention heads, and a hidden dimension size of $1024$. 
\\
\textbf{LLM.} We use the open-source Llama3.2 1B, 3B \cite{llama3.2}, and Llama3.1 8B as the LLM backbone. When LoRA \cite{hu2021lora} is applied with the LLM, the LoRA scaling factor hyperparameter is set to $16$. Following previous work \cite{G-Retriever}, we configure the LLM with a maximum input text length of $512$ and a maximum number of new tokens to generate of $32$.
\\
\textbf{Subgraph Construction.}  
We follow previous approaches \cite{G-Retriever} that select the top $k$ nodes and edges through subgraph construction by setting $k$ to $3$ for SceneGraphs dataset. For WebQSP dataset, $k = 3$ for nodes and $k = 5$ for edges. For the ExplaGraphs dataset, the entire graph fits within the LLM’s context window. Thus, setting $k$ to $0$ for retrieves the original graph without modification.
\\
\subsection{Implementation Settings (AGE AMAR)}\label{ImpSet}
When integrated with AMAR\cite{xu2025amar}, to fairly compare we keep the training settings of AMAR, setting the retrieved data to $100$ on WebQSP, Soft prompt length to 7, Beam search number to 8, and Max new tokens to 256 for the WebQSP dataset.
For the CWQ dataset, we set the retrieved data to 4, Soft prompt length to 16, Beam search number to 15, and Max new tokens to 256. With the Llama2 \cite{touvron2023llama} is trained with LoRA learning rate $5e-5$ scaling factor hyperparameter is set to $32$. 
%

\subsection{The choice of AGE architecture}\label{a}
\subsubsection{The choice of graph structure extractor architecture on AGE G-Retriever}
\begin{figure*}[ht]
\begin{center}
\includegraphics[width=0.98\linewidth]{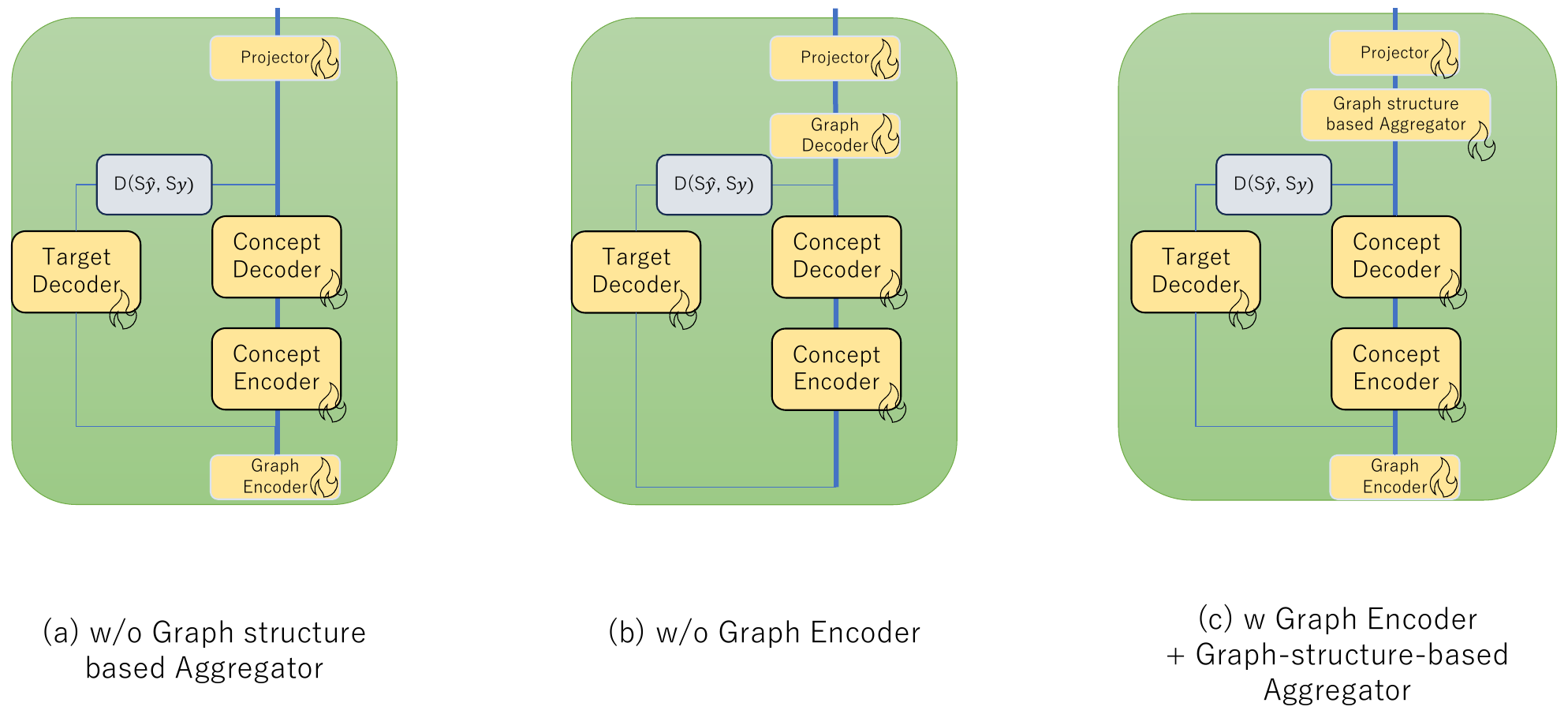}
\end{center}
   \caption{Investigation of core component arrangement: We tested our JEPA \cite{LeCun2022APT} architecture with three different GNN arrangements, including (a) graph encoder only, (b) graph-structure-based aggregator only, and (c) both of them.}
\label{tab:GNN_framework}
\end{figure*}
Figure \ref{tab:GNN_framework} shows the architectures with graph encoder only, graph-structure-based aggregator only, and both combined. The best-performing architecture is the combination of graph encoder and graph-structure-based aggregator. It achieves a Hit@1 score of $73.46$\% on the WebQSP dataset, improving upon $71.12$\% with Graph Encoder and $72.44$\% with graph-structure-based aggregator.
\subsubsection{The Choice of GNN on AGE G-Retriever}
\begin{table}[ht]
    \centering
    \begin{tabular}{lcc}
        \toprule
        \textbf{GNNs} & \textbf{WebQSP} & \textbf{ExplaGraphs} \\
        \midrule
        GCN               & 56.75 & 0.8321 \\
        GAT               & 61.42 & 0.8212 \\
        Graph Transformer & 62.53 & 0.8501 \\
        \bottomrule
    \end{tabular}
    \caption{Performance comparison of different GNNs on Llama3.2 1B.}
    \label{tab:graph_encoders}
\end{table}
In Table \ref{tab:graph_encoders}, our investigation extends to existing popular GNNs employed as both graph encoders and graph-structure-based Aggregator, including the Graph Convolutional Network (GCN) \cite{GCN}, Graph Attention Network (GAT) \cite{GraphAttention} and Graph Transformer \cite{GT}.
\\
This illustrates the performance comparison of these GNNs on the WebQSP and ExplaGraphs datasets.
On the WebQSP dataset, the GCN, GAT, and Graph Transformer achieve Hit@1 scores of $56.75$, $61.42$, and $62.53$, respectively. In the ExplaGraphs dataset, the Graph Transformer achieves the highest accuracy of $0.8501$, followed by the GCN with an accuracy of $0.8321$, and the GAT trailing slightly at $0.8212$.
\\
These findings emphasize the critical role of selecting an appropriate GNN architecture tailored to the unique properties and demands of each dataset. To maintain performance across various datasets, we choose the Graph Transformer for all experiments.
\subsubsection{The design of AGE with Generation Architecture}
As shown in Figure \ref{figureGA}, we provide more details of AGE with the randomly masked generative architecture (GA) introduced in Section 4.3. Designing AGE with GA aims to complement the input node to enhance the embedding of the graph-structure-based aggregator during the inference stage. To do this, we train the encoder-decoder with input nodes masked by a random mask at a masking ratio of $70$\%.Then, the encoder is trained to embed unmasked nodes, and the decoder reconstructs masked nodes through the target loss. On the other hand, prompt tuning loss is used to train the graph-structure-based aggregator and MLP for referring to edges $E^{*}$, and the MLP adjusts the aggregated embeddings to align with the LLM input dimension. During the inference stage, random masking is disabled. All input nodes are fed into the encoder-decoder to reconstruct the input for the graph-structure-based aggregator.
\begin{table}[ht]
\centering
\small
\caption{Analysis on the number of $GNN_{ge}$' layers with LLaMA 3.2 3B on WebQSP. GE refers to Graph Embedding.}
\begin{tabular}{cr|rr|rrr}
\toprule
& PT w/o GE
& \multicolumn{2}{c|}{G-Retriever}     
& \multicolumn{3}{c}{\makecell{AGE}}  \\

\# of layers & - & 2 & 4  & 1 & 2 & 4 \\
\midrule
Hit@1 ($\uparrow$) &      48.3 &   64.9   & 71.3  & 73.5 & 70.5  & 69.7     \\

Training time (Min./Epoch) ($\downarrow$) 
& 4.5  & 4.4  & 4.5 & 4.5 & 4.6  & 4.9 \\

Inference speed (Tokens/sec) ($\uparrow$) 
& 88.9     & 86.0    & 84.4   & 87.6 & 84.9  & 81.1    \\
\bottomrule
\end{tabular}
\label{tab:IoIOC}
\end{table}
\newline
\subsubsection{Analysis on the Layer Number of the Graph Encoder $GNN_{ge}$ on AGE G-Retriever}
Compared to G-Retriever, AGE's inference path has additional modules. While this might increase processing time, Table \ref{tab:IoIOC} indicates otherwise. For G-Retrievers, a deeper $GNN_{ge}$ performs better. In contrast, AGE performs better with fewer layers, as the added modules effectively substitute for reduced GNN layers. As a result, AGE achieves superior performance while maintaining the training time of the baseline method.
We provide further analysis on computational complexity in Appendix.
%
\subsubsection{Comparison with Deeper GNNs}
\begin{table}[ht]
\centering
\begin{tabular}{cccccc}
\toprule
 &  LLM 
 & GNN Layer 
 & Parameter
 & FLOPs (G)  
 & Acc \\
\midrule
G-Retriever  & Llama 3.2 1B        & 4          & 3.9 M      & 0.2 G    & 0.5595 \\
G-Retriever  & Llama 3.2 1B        & 20         & 11.3M      & 1.2 G    & 0.7238 \\
AGE G-Retriever & Llama 3.2 1B        & 2          & 7.8 M      & 1.1 G    & 0.8501 \\

\midrule

G-Retriever  & Llama 3.2 1B        & 4          & 3.9 M      & 0.2 G    & 0.7761 \\
G-Retriever  & Llama 3.2 1B        & 20         & 11.3M      & 1.2 G    & 0.8682 \\
AGE G-Retriever   & Llama 3.2 1B        & 2          & 7.8 M      & 1.1 G    & 0.9260 \\
\bottomrule
\end{tabular}
\caption{Compare AGE with DeeperGNN in ExplaGraphs test set}
\label{table:Deeper}
\end{table}

Table \ref{table:Deeper} compares the number of GNN layers and the performance of G-Retriever and Adaptive-masking for Graph Embedding models. G-Retriever with 20 layers is prepared as a model whose computational cost (GFLOPs) is similar to AGE with 2 layers.
\\
Applying the G-Retriever with $20$ layers largely improves performance. However, AGE still outperforms G-Retriever by approximately $10$ points when using Llama 3.2 1B and $6$ points when using Llama 3.2 3B, demonstrating AGE's superior performance.

\subsubsection{The choice of concept decoder and node sampler architecture.} 
\begin{table*}[t]
    \setlength{\tabcolsep}{4pt}
    \centering
    \small
    \begin{subtable}{0.45\linewidth}
        \centering
        \begin{tabular}{cccc}
            \toprule
            \makecell{Decoder \\ Depth} & \makecell{Para. \\ (M)} & \makecell{Expla \\ (Acc)} & \makecell{WebQSP \\ (Hit@1)} \\
            \midrule
            1 & 81 & 0.8501 & 62.5 \\
            2 & 85 & 0.7978 & 61.2 \\
            4 & 106 & 0.8123 & 57.4 \\
            \bottomrule
        \end{tabular}
        \caption{Concept Decoder}
        \label{table:Decoder}
    \end{subtable}
    \hspace{0.03\linewidth}
    \begin{subtable}{0.47\linewidth}
        \centering
        \begin{tabular}{cc|ccc}
            \toprule
            Depth & $d$ &\makecell{FLOPs \\ (G)} & \makecell{Expla \\ (Acc)} & \makecell{WebQSP \\ (Hit@1)} \\
            \midrule
            $1$ & $1024$ & 1.1 & 0.8501 & 62.5 \\
            $1$ & $2048$ & 1.6 & 0.8213 & 59.3 \\
            $2$ & $1024$ & 1.4 & 0.7906 & 63.1 \\
            \bottomrule
        \end{tabular}
        \caption{Node Sampler}
        \label{table:Maskselector}
    \end{subtable}
    \caption{Ablation studies for network architecture design.}    
\end{table*}

Table \ref{table:Decoder} illustrates our analysis of model performance across various concept decoder depths. We increased the decoder depth from 1 block to 4 blocks, thereby increasing the parameters from $81$M to $106$M. Despite this increase, the performance decreased, with scores dropping from $62.5$ to $57.4$ on WebQSP. The best performance is achieved with a decoder depth of 1 on both ExplaGraphs and WebQSP. Therefore, we choose a single transformer block to maintain the performance of the concept decoder in this work.
\\
As shown in Table \ref{table:Maskselector}, we investigate different network architectures for the node sampler design. Increasing the number of transformer blocks leads to marginal gains in performance on the WebQSP dataset, although it requires more memory. To maintain computational and performance efficiency, we selected a single transformer block with a hidden dimension of 1024 for the node sampler in all subsequent experiments.

\begin{figure*}[t]
\begin{center}
\includegraphics[width=0.9\linewidth]{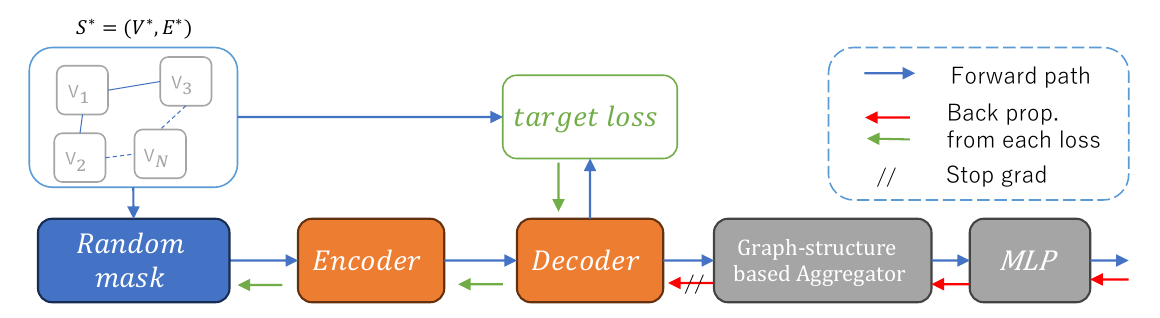}
\end{center}
   \caption{Adaptive-masking for Graph Embedding in Generation Architecture: The node embedding module is trained on both prompt tuning loss and target loss.}
\label{figureGA}
\end{figure*}
\subsubsection{The study on key nodes sampling}
\begin{figure*}[ht]
\centering
\small
\begin{minipage}{0.49\textwidth}
\centering
\begin{tikzpicture}
\begin{axis}[
xlabel={Sampling rate $\rho$},
ylabel={Accuracy},
xtick={0.3, 0.4, 0.5, 0.6, 0.7},
legend pos=south west,
ymin=0.70, ymax=0.87,
grid=major,
width=\textwidth,
height=6.5cm
]
\addplot[
color=blue,
mark=square,
]
coordinates {
(0.7,  0.7834)
(0.6,  0.7942)
(0.5,  0.8069)
(0.4,  0.8056)
(0.3,  0.8501)
(0.25, 0.8105)
};
\addlegendentry{Node Sampler}

\addplot[
color=red,
mark=triangle,
]
coordinates {
(0.7,  0.8149)
(0.6,  0.7743)
(0.5,  0.8176)
(0.4,  0.7797)
(0.3,  0.7815)
(0.25, 0.7761)
};
\addlegendentry{Page Rank}

\addplot[
color=green,
mark=triangle,
]
coordinates {
(0.7,  0.8086)
(0.6,  0.8068)
(0.5,  0.8000)
(0.4,  0.7942)
(0.3,  0.7996)
(0.25, 0.7896)
};
\addlegendentry{Degree Centrality}

\end{axis}
\end{tikzpicture}
\caption{Relationship of sampling rate with key node sampling strategy (on ExplaGraphs)}
\label{fig:keynode1}
\end{minipage}
\hfill
\begin{minipage}{0.49\textwidth}
\centering
\begin{tikzpicture}
\begin{axis}[
xlabel={Sampling rate $\rho$},
ylabel={Hit@1},
xtick={0.3, 0.4, 0.5, 0.6, 0.7},
legend pos=south east,
ymin=50, ymax=64,
grid=major,
width=\textwidth,
height=6.5cm
]
\addplot[
color=blue,
mark=x,
]
coordinates {
(0.7, 61.5479)
(0.6, 60.1351)
(0.5, 60.6880)
(0.4, 58.5381)
(0.3, 62.5307)
(0.25, 62.3464)
};
\addlegendentry{Node Sampler}

\addplot[
color=red,
mark=triangle,
]
coordinates {
(0.7,  61.67)
(0.6,  57.55)
(0.5,  56.14)
(0.4,  58.66)
(0.3,  55.65)
(0.25, 58.10)
};
\addlegendentry{Page Rank}

\addplot[
color=green,
mark=circle,
]
coordinates {
(0.7,  59.82)
(0.6,  57.06)
(0.5,  58.47)
(0.4,  58.72)
(0.3,  57.24)
(0.25, 56.94)
};
\addlegendentry{Degree Centrality}

\end{axis}
\end{tikzpicture}
\caption{Relationship of sampling rate with key node sampling strategy (on WebQSP)}
\label{fig:keynode2}
\end{minipage}
\end{figure*}

Figure \ref{fig:keynode1} and \ref{fig:keynode2} illustrates how the static strategy node sampler and RL-based node sampler performance in various sampling rate on ExplaGraphs and WebQSP, guiding our node sampler architecture setting.
\\
When using static PageRank \cite{Page1998PageRank} and Degree Centrality strategies for node sampling, higher sampling rates tend to better performance. However, this suggests that the key nodes that identified by these static methods are not sufficiently impactful. Lead to the Concept Encoder-Decoder needs a larger set of key nodes to effectively embed the graph, which then helps guide the LLM to produce the desired answers. 
\\
In contrast, RL-based node samplers can achieve high performance with lower sampling rates. This indicates that the key nodes chosen by RL-based methods more effectively support the Concept Encoder-Decoder, boosting the quality of the graph embedding. As a result, the LLM can produce the expected answers with fewer key nodes involved in the guidance process.
\subsubsection{The study on transferability}
\begin{table}[ht]
\centering
\begin{tabular}{llcc}
\toprule
Method 
& \makecell{\textbf{WebQSP} $\rightarrow$ \\ \textbf{Expla}} 
& \makecell{\textbf{Expla} $\rightarrow$ \\ \textbf{WebQSP}} \\
\\
\midrule
G-Retriever      & Llama 3.2 1B  & 0.5106 & 36.48 \\
\textbf{AGE G-Retriever}  & Llama 3.2 1B  & 0.5685 & 39.25 \\
G-Retriever      & Llama 3.2 3B  & 0.4404 & 50.35 \\
\textbf{AGE G-Retriever}  & Llama 3.2 3B  & 0.6021 & 53.53 \\
\bottomrule
\end{tabular}
\caption{Cross-Dataset Transfer Learning Performance.}
\label{table:transferability}
\end{table}

Table \ref{table:transferability} show the transferability of AGE when interacted with G-Retriever.
AGE support G-Retriever to strong transferability to transfer learned graph embedding encoding capabilities across datasets. When trained on a large dataset, AGE can enhance generation on a smaller dataset using the trained model. Notably, AGE trained on WebQSP on ExplaGraphs with Llama 3.2 3B outperforms transferability of GRAG.

\subsubsection{The study on number of retrieval}
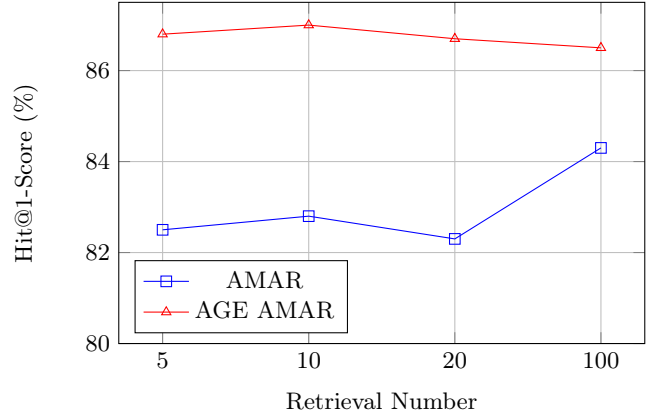
\begin{figure}[ht]
\centering
\begin{tikzpicture}
\begin{axis}[
xlabel={Retrieval Number},
ylabel={Hit@1-Score (\%)},
xtick={1, 2, 3, 4, 5, 6},
xticklabels={5, 10, 20, 100},
ymin=80, ymax=87.5,
legend pos=south west,
grid=major,
width=0.7\textwidth,
height=0.5\textwidth
]

\addplot[
color=blue,
mark=square,
]
coordinates {
(1, 82.5) (2, 82.8) (3, 82.3) (4, 84.3)
};
\addlegendentry{AMAR}

\addplot[
color=red,
mark=triangle,
]
coordinates {
(1, 86.8) (2, 87.0) (3, 86.7) (4, 86.5)
};
\addlegendentry{AGE AMAR}
\end{axis}
\end{tikzpicture}
\caption{Performance impart on vary number of retrieval on WebQSP}
\label{fig:impart_of_number_retrieval}
\end{figure}

As illustrated in Figure\ref{fig:impart_of_number_retrieval}, AMAR was designed to address the challenge posed by excessively long retrieved data inputs and to leverage rich information more effectively. However, when the volume of retrieved data is relatively small, AMAR’s performance shows minimal improvement, indicating that the recalled information is insufficient. In such cases, AGE is capable of mapping retrieved data to useful embeddings, eading to significant performance improvements, with scores of $86.8$ for 5 retrievals and $87.0$ for 10 retrievals.
\\
Conversely, when large amounts of data are retrieved, the accompanying noise complicates the ability of LLMs to identify and prioritize the most relevant information. AGE consistently maintains its performance with minimal variation, highlighting the robustness as $86.5$. Moreover,to fairly compare with AMAR, we choose 100  retrievals with an Hit1@ of $86.5$.
\subsubsection{Impart of AGE with LoRA Finetuning performance}

\begin{figure*}[ht]
\begin{center}
\includegraphics[width=0.95\linewidth]{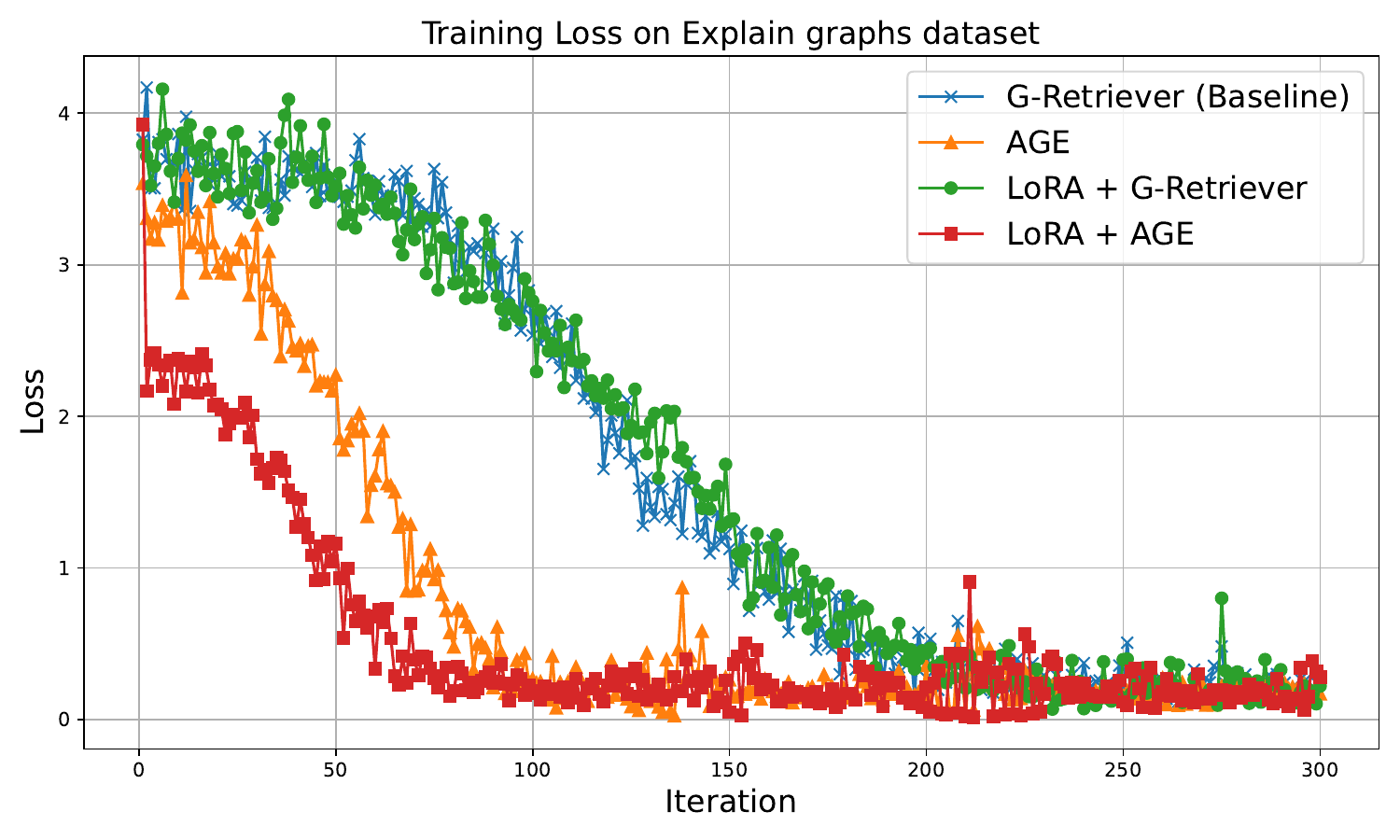}
\end{center}
   \caption{Training loss on Explain Graphs.}\label{fig:explain_impart_of_training_epoch}
\end{figure*}

\begin{figure*}[ht]
\begin{center}
\includegraphics[width=0.95\linewidth]{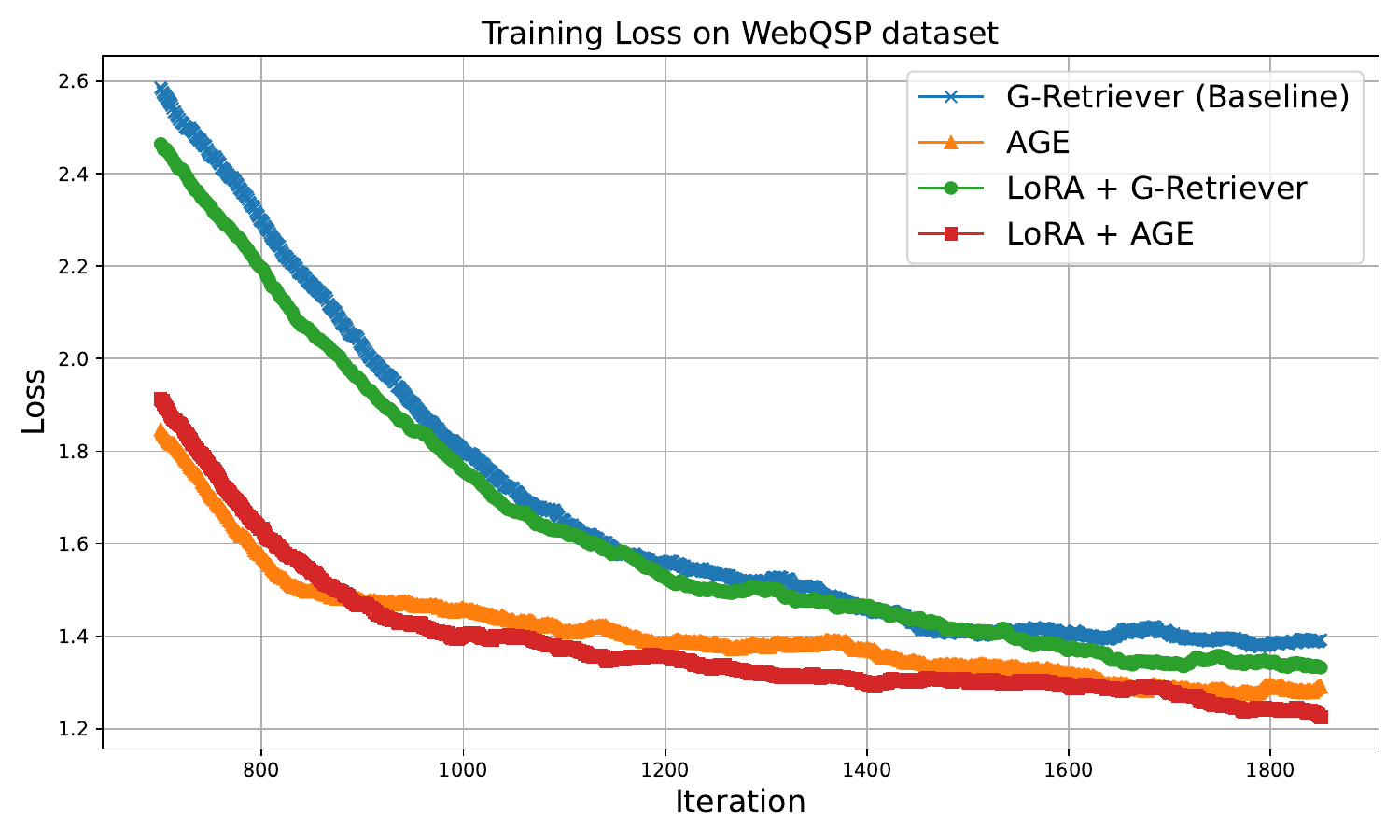}
\end{center}
   \caption{Training loss on WebQSP}\label{fig:web_impart_of_training_epoch}
\end{figure*}

We conduct extensive experiments to compare the trend of training loss with G-Retriever baselines, as illustrated in Figure\ref{fig:explain_impart_of_training_epoch} and \ref{fig:web_impart_of_training_epoch}. AGE provides structural constraints in the embedding space through mask-based SSL with reconstruction objective, enabling LLMs to better separate signals and capture relationships within it, leading to increased convergence rate and lower loss observed in the initial training stage compared with the G-Retriever.

\subsubsection{The choice of RL method for Node Sampler}

\begin{table*}[]
\centering
\small
\caption{The study of sampling strategy on Node Sampler (with Llama3.2 1b on PEFT, on ExplaGraphs and WebQSP).}
\begin{tabular}{cccc}
\toprule
& Gumbel-Softmax     & Straight-Through (ST) Estimator    & REINFROCE     \\
\midrule
ExplaGraphs       & 0.8378        & 0.8241   & 0.8501       \\ 
WebQSP            & 68.5          & 66.7     & 69.1         \\ 
\bottomrule
\end{tabular}
\label{table:RLchoicemethod}
\end{table*}

Table \ref{table:RLchoicemethod} show Both Gumbel-Softmax and REINFORCE demonstrate strong performance on the ExplaGraphs dataset, outperforming the Straight-Through Estimator. On the WebQSP dataset, REINFORCE leads slightly, indicating it may be the most effective method among the three tested for this task. From these observations, we decided to use REINFROCE through the experiments.

\subsubsection{The chose different input features for LLMs during training and inference}

\begin{table*}[]
\centering
\small
\caption{Performance of AGE in PEFT (with Llama3.2 1b , on ExplaGraphs and WebQSP).}
\begin{tabular}{cccc}
\toprule
& \makecell{PEFT G-Retriever}    
& \makecell{AGE \\ w $h_{target}$ as LLM input}   
& \makecell{AGE \\ w $h_{out}$ as LLM input}    \\
\midrule
ExplaGraphs       & 0.7328   & 0.8212   & 0.8501       \\ 
WebQSP            & 65.3     & 66.7     & 69.1         \\ 
\bottomrule
\end{tabular}
\label{table:outputperformance}
\end{table*}
Table \ref{table:outputperformance} show AGE with $h_{out}$ as LLMs input in inference state outperforms AGE with $h_{target}$ as input on both datasets. This indicates that representing nodes using a concept encoder-decoder $h_{out}$ is more effective than the target encoder in downstream LLMs input tasks. Based on these observations, we concluded that designing the connection of $h_{target}$ during training and $h_{out}$ during inference to the downstream LLM not internalizes the learning-inference mismatch. Instead, it allows the student model has already surpassed the performance of the teacher model, allowing for a more robust representation.

\subsubsection{The stability of Target encoder}
\begin{table*}[]
\centering
\small
\caption{Performance of Target Encoder on ExplaGraphs, trained with PEFT using Llama3.2 1b.}
\begin{tabular}{ccccc}
\toprule
Loss type
& w/o norm + w/o EMA   
& \makecell{norm}
& \makecell{EMA}    
& norm + EMA  
\\
\midrule
L1           &  0.7978    &  0.8375     &  0.8194   & 0.8303            \\ 
MSE          &  0.8357    &  0.8501     &  0.8375   & 0.8501            \\ 
\bottomrule
\end{tabular}
\label{table:targetsability}
\end{table*}
Due to differences in node representations during the training inference stage, using identical parameters for both the concept encoder and the target encoder helps prevent distributional shifts. We apply two popular techniques to enhance the stability provided by the target encoder for the concept encoder-decoder during the training stage.
EMA weights are defined as an exponential moving average of the encoder weights, and normalization is applied to enhance stability during the learning process. Normalization ensures consistent activation distributions and reduces internal covariate shift.
Table \ref{table:targetsability} shows that MSE performs better with L1, and normalization alone achieved the highest score. These results indicate that normalization consistently improves encoder stability and performance, and adding EMA offers further enhancements.
%

\subsubsection{The landscape of existing KGQA methods}
\begin{figure*}[ht]
\begin{center}
\includegraphics[width=0.98\linewidth]{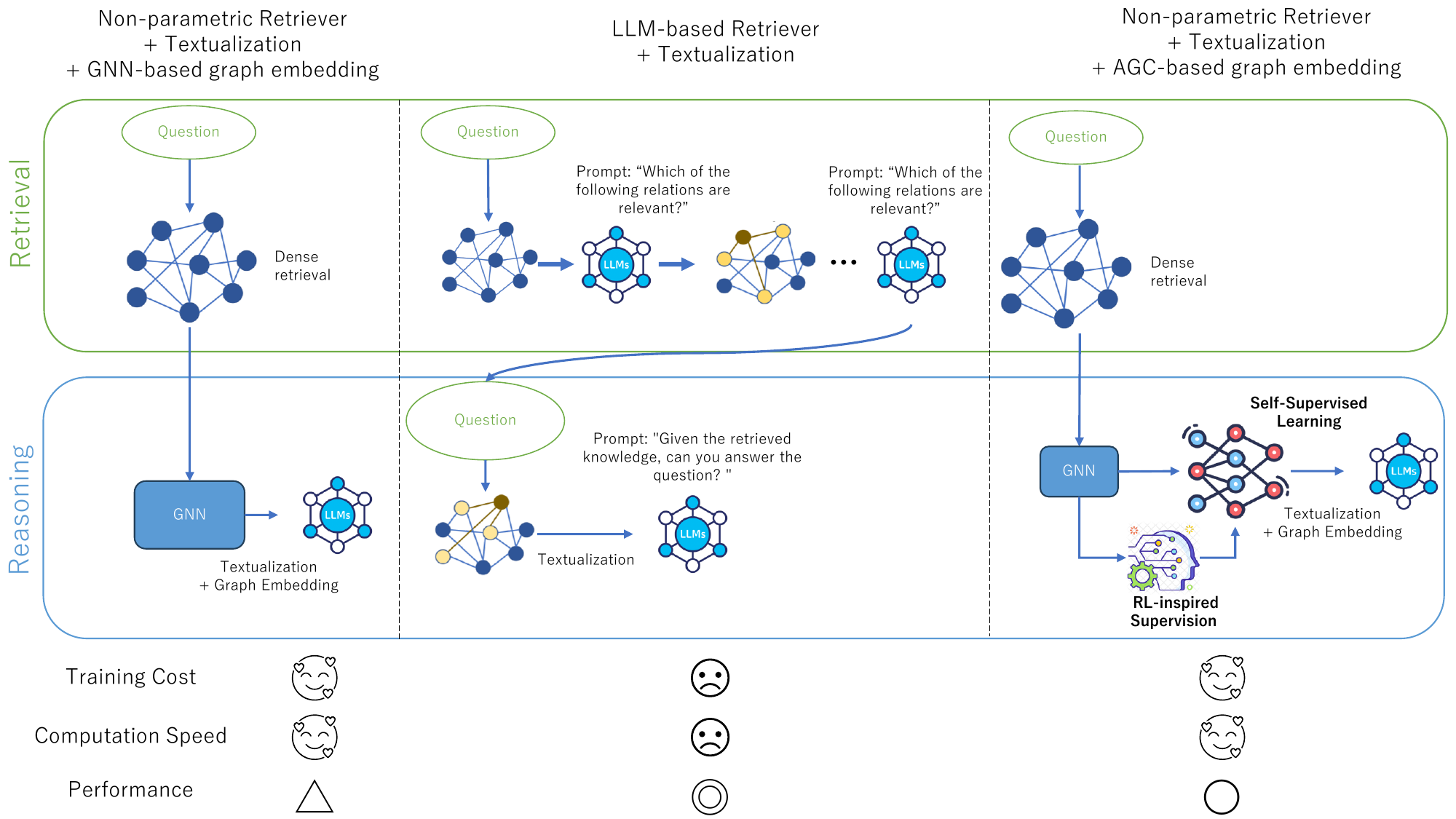}
\end{center}
   \caption{The landscape of existing KGQA methods. GNN-based methods reason on dense subgraphs as they can handle complex and graph information. LLM-based methods employ the same LLM for both retrieval and reasoning due to its ability to understand natural language.}\label{landscape_method}
\end{figure*}
Figure \ref{landscape_method} illustrates the spectrum of current Knowledge Graph Question Answering (KGQA) approaches regarding KG retrieval and reasoning capabilities. Graph Neural Network (GNN)-based methods, including NSM \cite{NSW}, ReaRev \cite{mavromatis-karypis-2022-rearev}, and G-Retriever \cite{G-Retriever}, perform reasoning on retrieved dense subgraphs by utilizing the GNN to embed graph structures.
\\
Recent LLM-based methods leverage the power of LLMs for both retrieval and reasoning. ToG \cite{think_on_graph} uses the LLM to retrieve relevant facts hop-by-hop. RoG \cite{reasoning_on_graphs} uses the LLM to generate plausible relation paths which are then mapped on the KG to retrieve the relevant information. However, the frequent calls to the LLM significantly increase the training and inference costs.
\\
In this work, we improve LLM reasoning by enhancing the graph embedding of the GNN method with RL-inspired supervision integrated into the SSL framework. This improves the performance of the non-parametric retriever to levels comparable to those of LLM-based retrievers.

\subsection{Additional Qualitative Evaluation}\label{QE}
We provide additional visualizations in Figures \ref{MoreQualitativeEval_A1}, \ref{MoreQualitativeEval_A2} on the ExplaGraphss dataset and Figure \ref{MoreQualitativeEval_A3} on WebQSP dataset.
\\
In the first row of Figure \ref{MoreQualitativeEval_A1}, we consider the addition of node text information and its visualization.
It is easier to infer the auxiliary node embeddings "Payday loans" and "For the disadvantaged" from an key node embeddings "Provide assistance". Conversely, it is more challenging to infer the key node embeddings "Provide assistance" from the auxiliary node embeddings "help society" and "available".
\\
Similarly, it is easier to infer "Bullying, However they like, Banned" mask node embeddings from a "Expensive clothes, Students" key node embeddings. Conversely, it is more challenging to infer a "Expensive clothes, Students" mask node embeddings from a "Bullying, However they like, Banned" key node embeddings.
\\
In the second row of Figure \ref{MoreQualitativeEval_A1}, when considering the addition of node text information and its visualization, it is easier to infer the auxiliary node embeddings "Motivation" and "Students work harder" from the key node embedding "Student loans". Conversely, it is more challenging to infer a "Student loans" auxiliary node embeddings from "Motivation" and "Students work harder" key node embeddings. Similar things are shown in Figure \ref{MoreQualitativeEval_A2} and Figure \ref{MoreQualitativeEval_A3}.
\begin{figure*}[t]
\begin{center}
\includegraphics[width=0.95\linewidth]{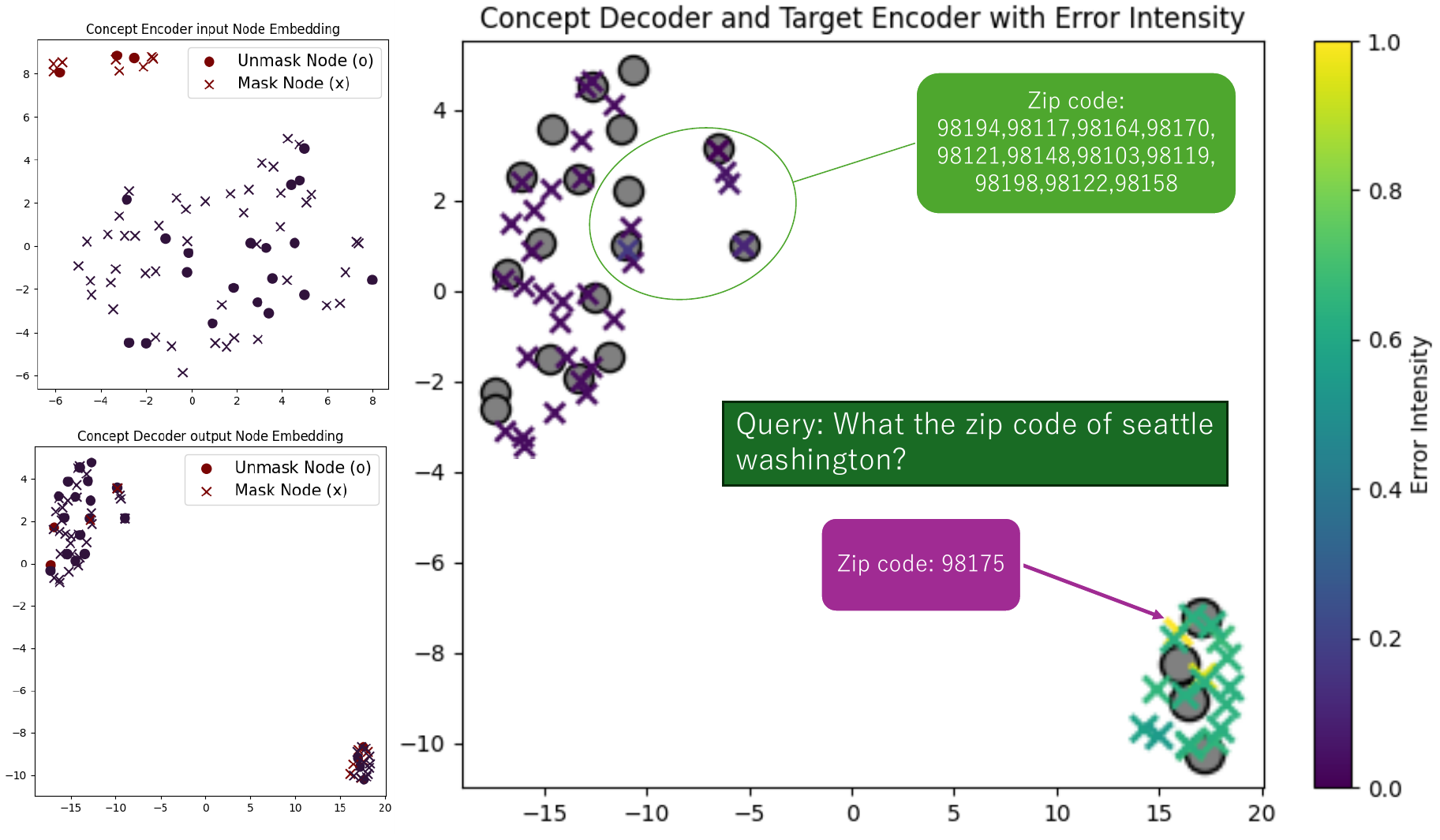}
\end{center}
   \caption{Failure case visualization of AGE on WebQSP dataset.}
\label{fig:failcase}
\end{figure*}
\textbf{Failure Case Analysis:} Furthermore, we provide a failure case on the WebQSP dataset where AGE was trained with LLaMA 3.2 1B using a sampling rate of $0.3$. In this case, the query is "What is the zip code of Seattle, Washington?".  Based on the query, the retriever is provided with the node "98175," which is one of the true answers. However, the response of the LLM lacks this node. To analyze this, we visualized the node sampling results from the concept encoder's input and the concept decoder's output, as shown in Figure \ref{fig:failcase}.
In the first column (top row), similar to the above visualization, the graph encoder process maintains the clustering structure of text graph embeddings to provide input to the concept encoder.
Additionally, the output of the concept encoder-decoder module (bottom row) shuffles the colored nodes.
In the second column, the target loss of each auxiliary node is represented using a color bar, where 1.0 indicates the maximum error in the test set and 0.0 the minimum error. In this column, the top left side nodes "98194, 98117, 98164,..." include low target loss auxiliary nodes, and the key nodes are the true answers. In parallel, the bottom right side node "98175" is an auxiliary node with a high target loss. This may be why the LLM omits this node in its response, and adjusting the trainable sampling rate could be a solution.
\begin{table}[tb]
\centering
\begin{tabular}{ll|cc}
\toprule
\textbf{Method} 
& \textbf{LLM} 
& \textbf{Hit@1} 
&  \textbf{Training} \textbf{Time}  (min/epoch) 
\\
\midrule

 
\midrule
 G-Retriever      & Llama 2 7B & 70.5 & 6.2 \\
 AGE G-Retriever  & Llama 3.2 1B       & 62.5 & 2.0 \\
 AGE G-Retriever  & Llama 3.2 3B       & 73.5 & 4.5  \\
 AGE G-Retriever  & Llama 3.2 8B       & 78.3 & 6.4  \\
\midrule
 G-Retriever       & Llama 2 7B LoRA        & 73.8  & 6.9 \\
 AGE G-Retriever   & Llama 3.2 1B LoRA      & 69.1  & 2.4 \\
 AGE G-Retriever   & Llama 3.2 3B LoRA      & 77.3  & 5.9 \\
 AGE G-Retriever   & Llama 3.2 8B LoRA      & 80.3  & 7.3 \\

 AGE AMAR          & Llama 2 7B LoRA        & 86.5   & 8.7 \\
\bottomrule
\end{tabular}
\caption{Training cost of AGE G-Retriever on the WebQSP dataset.}
\label{tab:train_speed}
\end{table}
\begin{table}[]
\centering
\small
\begin{tabular}{c|c||ccc}
\toprule
& \textbf{G-Retriver} & \textbf{AGE} & \textbf{AGE} & \textbf{AGE}\\
\midrule
\textbf{LLM size} & \textbf{Llama 2 7b}  
& \textbf{Llama 3.1 8b} & \textbf{Llama 3.2 3b}  & \textbf{Llama 3.2 1b}\\
\midrule
\textbf{\makecell{All Parameters \\ (B)}}   
& 6.8        & 8.1        & 3.3         & 1.3   \\
\midrule
\textbf{\makecell{Trainable \\ Para (B)}}
& 0.041  & 0.087      & 0.078  & 0.072 \\
\midrule
\textbf{\makecell{Inference speed \\ (Tokens/sec.)}}
& 97.0   &81.4        &87.6    &148.5 \\
\midrule
\textbf{Hit@1}                & 70.49      & 78.25       & 73.46       & 62.53     \\
\bottomrule
\end{tabular}
\caption{Inference speed of AGE on the WebQSP dataset.}
\label{tab:val_speed}
\end{table}
\subsection{Discussion on the Complexity}\label{computationcost}
\subsubsection{Training Computational Resources on AGE G-Retriever}
 Following the previous G-Retriever \cite{G-Retriever} method, we utilized the same two A100 GPUs, each with 80GB of memory, and conducted tests on the Llama3-8b, Llama3.1-1B, and Llama3.1-3B on WebQSP datasets. Our experiments had a training batch size of 16 and an evaluation batch size of 32, yielding the following results in Table \ref{tab:train_speed} for training cost and Table \ref{tab:val_speed} for validation speed.
\\
The Table \ref{tab:train_speed} shows the training speed and performance of AGE on the WebQSP dataset. The PEFT setting, without the graph RAG component, takes 18.7 min/epoch through prompt tuning and 19.0 min/epoch when applied with LoRA. Subsequently, the G-Retriever approach via graph RAG reduces graph size and speeds up training time. 
\\
By enhancing the embedding module on the graph RAG component, AGE with Llama3.1 8B achieves a higher Hit@1 of $78.25$ in $6.4$ minutes per epoch. In the tuned LLM setting, AGE with Llama3.1 8B and LoRA achieves a Hit@1 of $80.34$ in $7.3$ minutes per epoch. These results highlight that AGE with Llama3.2 3B outperforms G-Retriever with Llama2 7B, achieving better performance without longer training time.

\subsubsection{Inference Computational Resources on AGE G-Retriever}
\begin{table}[tb]
\centering
\scriptsize
\resizebox{\linewidth}{!}{%
\begin{tabular}{llc ccc c}
\toprule
& LLM & \makecell{Non-parameter \\ Retriever} & \multicolumn{2}{c}{\makecell{Trainable \\ Retriever}} & WebQSP & CWQ \\
&  & & GNN & LLM & Hit@1 & Hit@1 \\
\midrule
ToG                  & Llama2-70B   &  &  & \checkmark & 68.9 & 57.6        \\
RoG              & Llama2-7B    &  &  & \checkmark & 74.2 & 56.4        \\
ReKnoS             & Llama3.1-8B  &  & \checkmark  & \checkmark & 67.9 &  56.7       \\
DualR                    & Llama2-13B   &  & \checkmark  & \checkmark & 78.3 &  58.0       \\
\midrule
StructGPT                   & ChatGPT     &  &  & \checkmark & 72.6 &   55.3      \\
ToG                   & ChatGPT     &  &  & \checkmark & - & 76.2 \\
ToG-2               & ChatGPT     &  &  & \checkmark & 81.1 &   -     \\
RoG              & ChatGPT     &  &  & \checkmark & - & 80.0 \\
ReKnoS             & ChatGPT     &  & \checkmark & \checkmark & 81.1 &    58.5   \\
GNN-RAG & ChatGPT   &  & \checkmark &  & 85.7 &   66.8      \\
PoG                     & ChatGPT     &  &  & \checkmark & - & 82.0 \\
DualR                     & ChatGPT     &  & \checkmark & \checkmark & - & 82.8 \\
\midrule
KBQA      & GPT-4       &  &  & \checkmark & 72.5 &  -       \\
ReKnoS              & GPT-4       &  & \checkmark & \checkmark & 84.9 &   68.2      \\
ToG                    & GPT-4       &  &  & \checkmark & 82.6 & 69.5      \\
PoG                     & GPT-4       &  &  & \checkmark & 87.3 & 75.0 \\
DualR                     & GPT-4       &  & \checkmark & \checkmark & 87.6 &  73.6        \\
\midrule
GraphToken                 & Llama2-7B        & \checkmark &  &  & 57.1 &  -       \\
G-Retriever               & Llama2-7B-LoRA   & \checkmark &  &  & 70.2 &  -       \\
\textbf{AGE G-Retriever}                    & Llama3.1 8B-LoRA & \checkmark &  &  & 80.3 &  -       \\

AMAR                       & Llama2-7B-LoRA  & \checkmark &  &  & 84.3 &  82.9    \\
AMAR                       & Llama2-13B-LoRA  & \checkmark &  &  & 83.3 &  83.1    \\
\textbf{AGE AMAR}                           & Llama2-7B-LoRA  & \checkmark &  &   & 86.5 &  85.2    \\
\textbf{AGE AMAR}                           & Llama2-13B-LoRA  & \checkmark &  &  & 86.2    &  85.1    \\
\bottomrule
\end{tabular}%
}
\caption{Performance comparison of trainable retriever with AGE.}
\label{tab:trainableretriever}
\end{table}
Table \ref{tab:val_speed} presents the validation speed and performance of various AGE configurations on the WebQSP dataset. Among the AGE models, Llama 3.2 3B model offers a balanced performance with a Hit@1 and an inference speed of $87.6$ tokens per second. The AGE with Llama 3.2 1B achieves a significantly higher inference speed of $148.5$ tokens per second while maintaining a lower Hit@1. This increased speed can be attributed to the reduced number of parameters in the 1B model, which allows for faster computation and more efficient processing, albeit at the expense of some accuracy. 
\\
These results indicate that while higher parameter models like AGE+Llama 3.1 8B provide superior accuracy, lower parameter models such as AGE+Llama 3.2 1B offer significantly increased processing speeds, supporting diverse application requirements.
\subsubsection{Comparison with trainable retriever methods}
AGE, utilizing a non-parametric retriever, achieves accuracy levels comparable to state-of-the-art models that employ trainable parametric retrievers. As shown in Table \ref{tab:trainableretriever}, AGE (Llama3.1 8B-LoRA with a non-parametric retriever) attains a Hit@1 score of $80.3\%$ on the WebQSP dataset, closely approaching DualR (ChatGPT with a parametric retriever), which achieves $82.8\%$. This demonstrates that AGE effectively bridges the performance gap between non-parametric and parametric retriever models, achieving high accuracy without the additional complexity and training overhead associated with parametric retrievers. This performance notably surpasses other models employing non-parametric retrievers, such as GraphToken (Llama2-7B) with $57.1\%$ and G-Retriever (Llama2-7B-LoRA) with $70.2\%$. 
\\
The substantial increase in accuracy demonstrates that AGE enhances reasoning capabilities without relying on trainable parametric retrievers. This positions AGE as a leading approach within non-parametric retriever frameworks, closing the performance gap with models that utilize more complex and resource-intensive trainable retrievers. AGE can be deployed to train and perform inference on two RTX 2080Ti 11GB GPUs or one A100 80GB GPU.
\begin{figure*}[]
    \centering
    \includegraphics[width=0.98\linewidth]{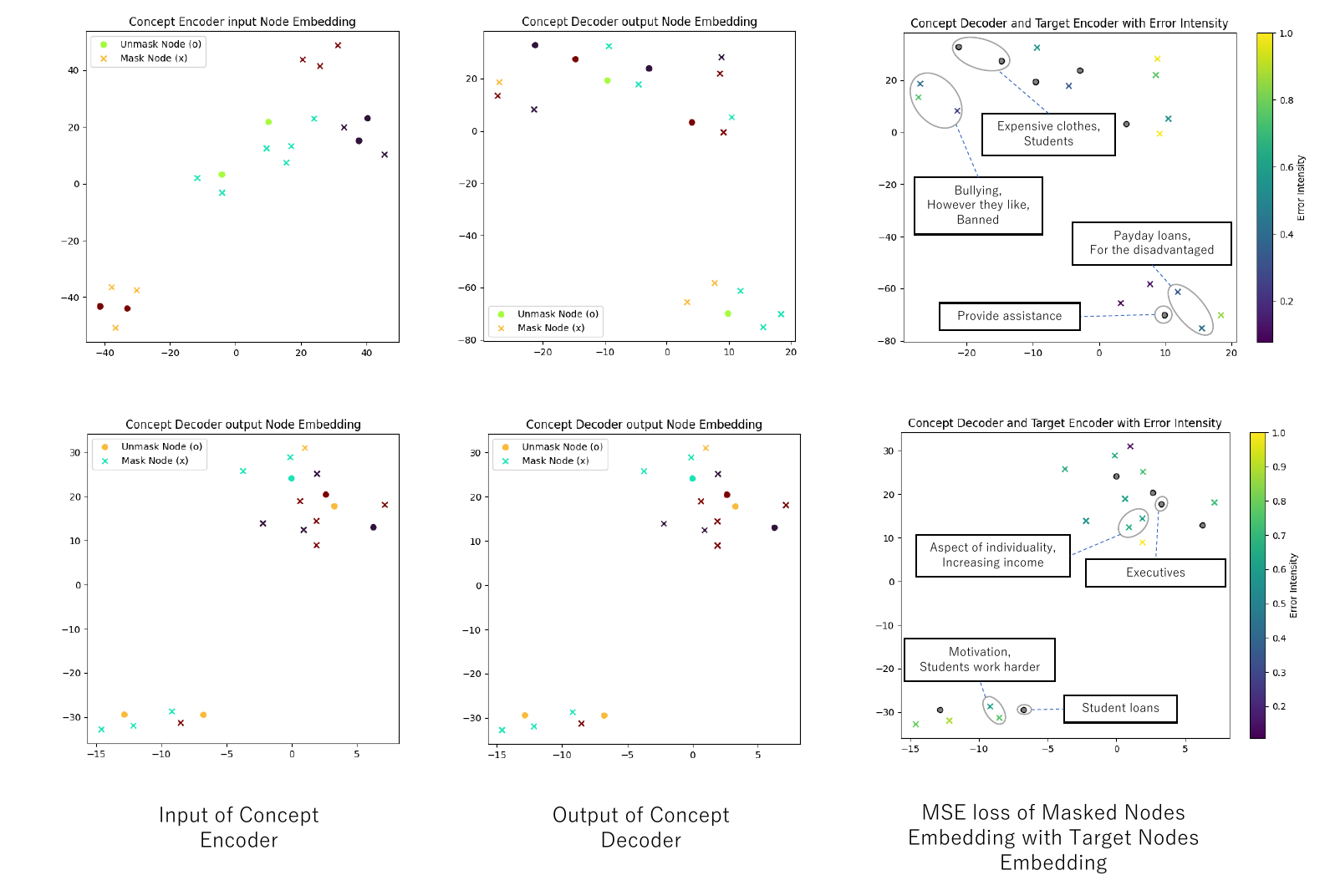}
    \caption{An example visualization of AGE on the ExplaGraphs dataset.}
    \label{MoreQualitativeEval_A1}
\end{figure*}
\begin{figure*}[]
    \centering
    \includegraphics[width=0.98\linewidth]{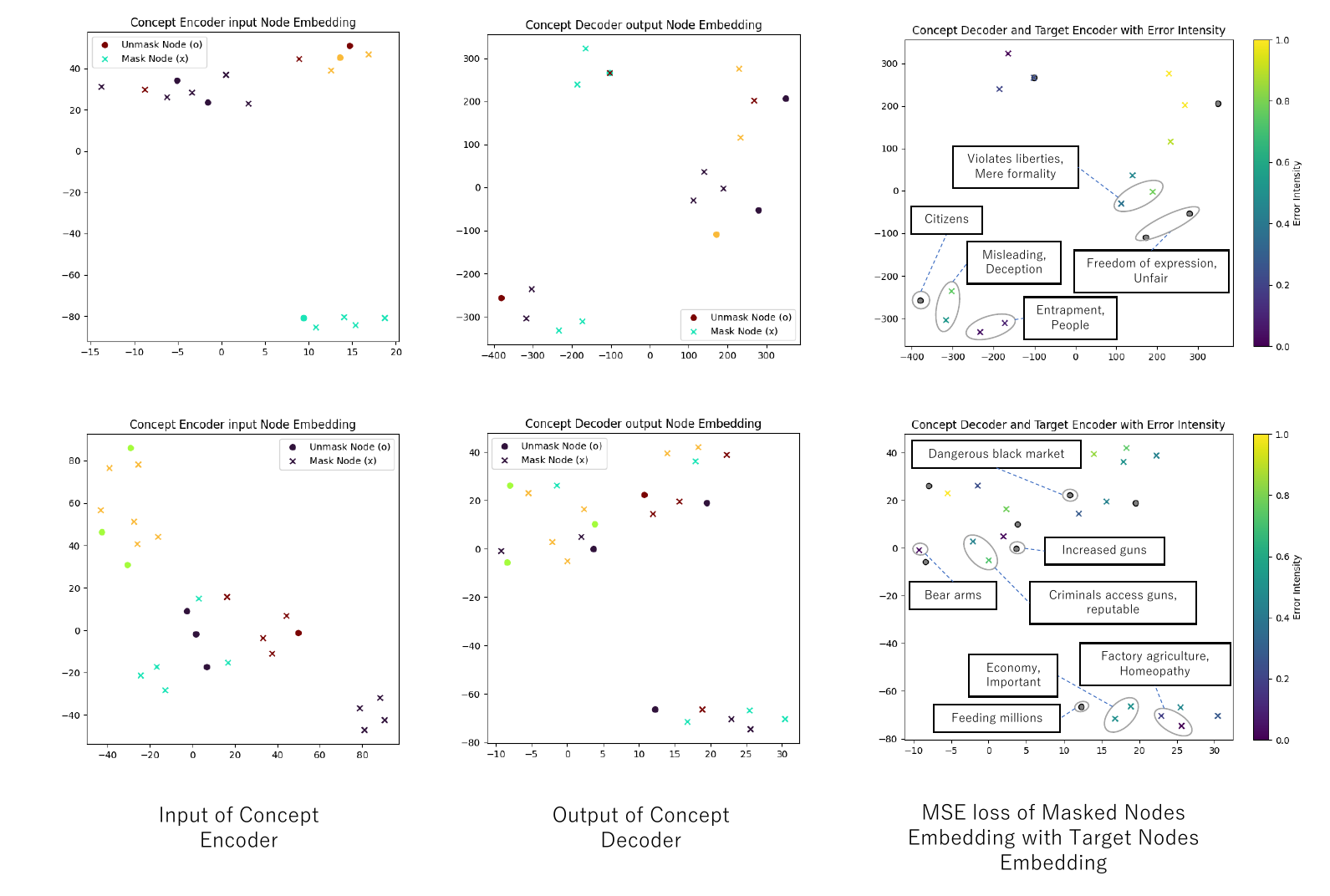}
    \caption{Another example visualization of AGE on the ExplaGraphs dataset.}
    \label{MoreQualitativeEval_A2}
\end{figure*}
\begin{figure*}[]
    \centering
    \includegraphics[width=0.98\linewidth]{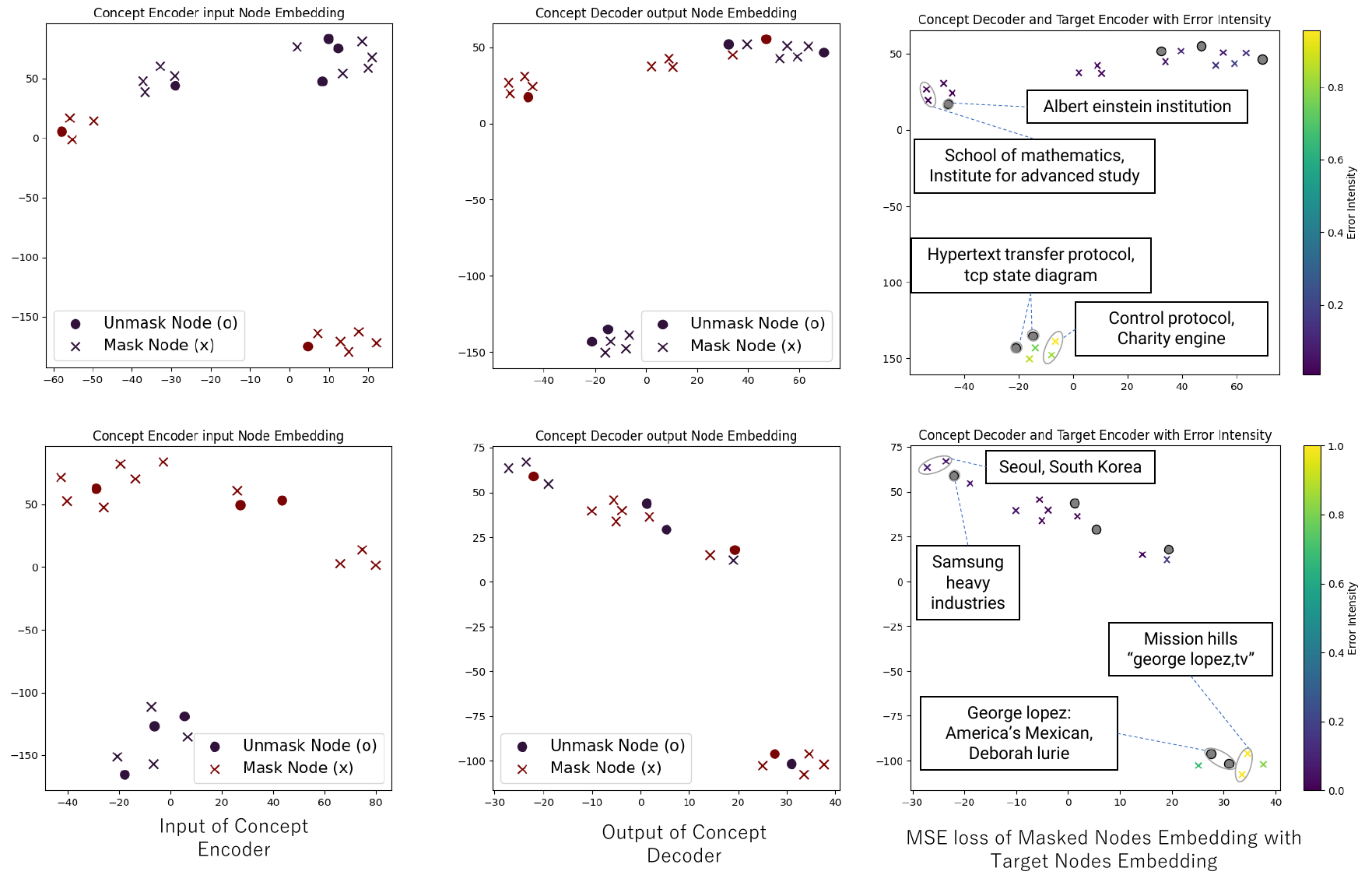}
    \caption{An example visualization of AGE on the WebQSP dataset.}
    \label{MoreQualitativeEval_A3}
\end{figure*}

\end{document}